\documentclass[a4,aps,amsmath,floatfix,nofootinbib,longbibliography,twocolumn,prb,superscriptaddress,prb]{revtex4-1}

\usepackage{graphicx}
\usepackage[colorlinks=true,linkcolor=red]{hyperref}
\usepackage{xcolor}
\usepackage{dcolumn}
\usepackage{bm} 
\usepackage{amsmath}
\usepackage{amssymb}
\usepackage{csquotes}

\newcommand{\pcsadd}{Center for Theoretical Physics of Complex Systems, Institute for Basic Science(IBS), Daejeon 34126, Korea}
\newcommand{\ustadd}{Basic Science Program(IBS School), Korea University of Science and Technology(UST), Daejeon 34113, Korea}

\newcommand{\mh}{\mathcal{H}}
\newcommand{\vq}{\vec{q}}
\newcommand{\vpsi}{\vec{\psi}}
\newcommand{\ve}{\vec{e}}
\newcommand{\vk}{\vec{k}}

\begin{document}

\title{Microwave photonic crystals as an experimental realization of a combined honeycomb-kagome lattice}

\author{Wulayimu Maimaiti}
\affiliation{\pcsadd}
\affiliation{\ustadd}

\author{Barbara Dietz}
\email{dietz@lzu.edu.cn}
\affiliation{School of Physical Science and Technology, and Key Laboratory for Magnetism and Magnetic Materials of MOE, Lanzhou University, Lanzhou, China}

\author{Alexei Andreanov}
\email{aalexei@ibs.re.kr}
\affiliation{\pcsadd}
\affiliation{\ustadd}

\date{\today}

\begin{abstract}
In 2015 experiments were performed with superconducting microwave photonic crystals emulating artificial graphene [B. Dietz {\it et al.}, Phys. Rev. B {\bf 91}, 035411 (2015)]. The associated density of states comprises two Dirac points with adjacent bands including van Hove singularities, thus exhibiting the characteristic features originating from the extraordinary electronic band structure of graphene. They are separated by a narrow region of particularly high resonance density corresponding to a nearly flatband in the band structure, which is reminiscent of that of a  honome lattice -- a combination of two sublattices: honeycomb and kagome. We demonstrate that, indeed, the density of states, and also the eigenmode properties and the fluctuations in the resonance-frequency spectra are well reproduced by a tight-binding model based on the honome lattice. A good description was achieved by means of the reverse Monte-Carlo approach, thereby confirming our intepretation of the microwave photonic crystal as an experimental realization of a honome lattice and providing an answer to longstanding problem, namely the understanding of the origin of the flatband bordered by two Dirac points, generally observed in microwave photonic crystals of different shapes.
\end{abstract}


\maketitle

\section{Introduction}
\label{sec:intro_motivation}

The extraordinary band structure of graphene, a monolayer of carbon atoms on a hexagonal lattice triggered numerous extensive theoretical~\cite{beenakker2008colloquium,castro2009the} and experimental studies.~\cite{novoselov2004electric,ponomarenko2008chaotic} Particularly the linear dispersion relation around the touch points of the conduction and valence bands -- commonly referred to as Dirac points (DPs) -- results in relativistic phenomena occurring in graphene.~\cite{castro2009the,geim2007the,avouris2007carbon,miao2007phase,beenakker2008colloquium,abergel2010properties} These features result from the symmetry properties of the honeycomb structure which is formed by two interpenetrating triangular lattices. Consequently, photons (bosons) or waves propagating in a spatially periodic potential with a honeycomb structure may comprise in their energy spectra regions where they are effectively described by the Dirac equation for spin-1/2 fermions. By now there exist numerous realizations of artificial graphene~\cite{polini2013artificial} using two-dimensional electron gases exposed to a honeycomb potential lattice,~\cite{singha2011two,nadvornik2012from} molecular assemblies arranged on a copper surface,~\cite{gomes2012designer} ultracold atoms in optical lattices~\cite{tarruell2012creating,uehlinger2013artificial} and photonic crystals.~\cite{parimi2004negative,raghu2008analogs,joannopoulos2008photonic,bittner2010observation,kuhl2010dirac,sadurni2010playing,bittner2012extremal,bellec2013topological,bellec2013tight,rechtsman2013strain,rechtsman2013topological,khanikaev2013photonic} 

Particularly, graphene-like lattice structures have been realized in flat microwave resonators,~\cite{stoeckmann1990quantum,sridhar1991experimental,graf1992distribution} called Dirac billiards because their resonance spectra exhibit Dirac points.~\cite{dietz2013lifshitz,dietz2015spectral,dietz2019from} It was demonstrated in Ref.~\onlinecite{dietz2015spectral} that for the bands framing the lower Dirac point the fluctuation properties of the resonance frequencies can be described by a honeycomb-lattice based tight-binding model (TBM) which comprises 1st, 2nd and 3rd nearest-neighbor (n.n.) hoppings and takes into account wave function overlaps between neighboring lattice sites.~\cite{reich2002tight} Spectral properties were extensively studied in rectangular superconducting microwave Dirac billiards in Ref.~\onlinecite{dietz2015spectral}. The resonance spectrum comprises two DPs and a narrow frequency range of exceptionally high resonance density between them associated with a nearly flatband (FB) in the band structure, of which the origin has not been understood up to now. Application of a honeycomb-lattice based TBM yielded a very good description for the bands below the FB, whereas above the FB a closed tight-binding interpretation was no longer possible. Furthermore, the occurrence of the FB in the band structure could not be explained by such a TBM. In Ref.~\onlinecite{jacqmin2014direct} a similar band structure containing two bands framing a Dirac point in the lowest energy range and an adjacent FB was observed for a two-dimensional lattice of coupled micropillars which were etched into a semiconductor microcavity. In the present article, we introduce a TBM which is capable of describing the experimental spectrum from the lower band edge below the first DP up to the upper one above the second DP. The underlying lattice is formed by two sublattices, honeycomb and kagome, and was named super-honeycomb lattice in Refs.~\onlinecite{jacqmin2014direct,lan2012coexistence,lu2017two,zhong2017transport}. For brevity, we call it honome lattice. The model provides a description of the spectrum including {\it both} DPs, the bands framing them and, in addition, accounts for the nearly FB.

The article is organized as follows. Section~\ref{sec:intro} summarizes the salient features of the microwave photonic crystal, Sec.~\ref{sec:TBM} introduces the tight-binding description of its spectrum in the bands below the FB and Sec.~\ref{sec:model} presents the improved TBM which is based on the honome lattice. The fitting procedure is described in Sec.~\ref{sec:methods} and detailed in Sec.~\ref{sec:results}. In Sec.~\ref{Comparison}  we compare the wave-functions and the fluctuation properties in the eigenfrequency spectra of the honome TBM matrix to experimental and numerical results, obtained by solving the associated Helmholtz equation. Finally, we summarize and discuss the results in Sec.~\ref{sec:conclusion}.

\section{Photonic crystals and Dirac billiards}
\label{sec:intro}

\begin{figure}[h]
    \centering
    \includegraphics[width=0.9\columnwidth]{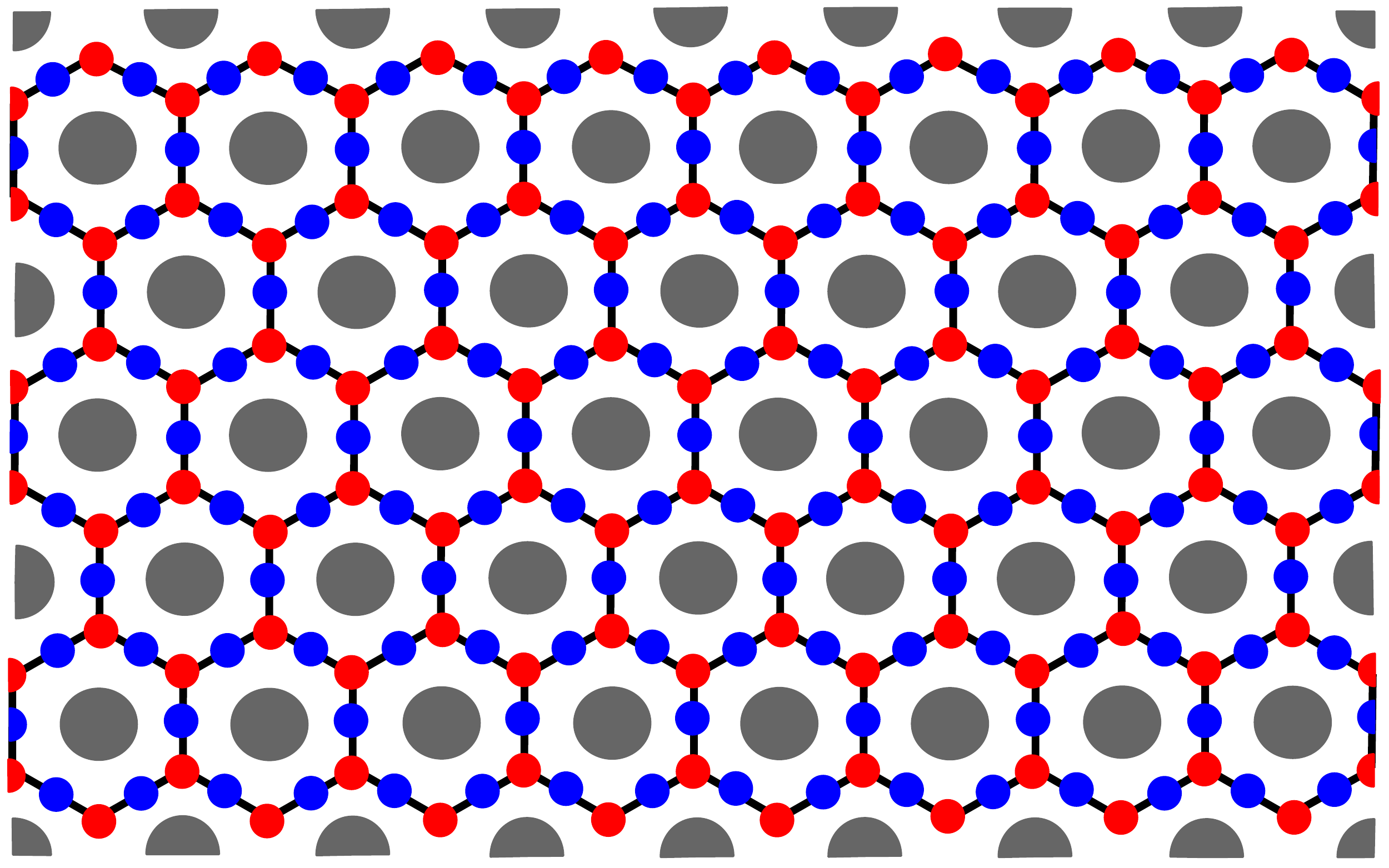}
    \caption{Schematic view of the triangular lattice structure of the billiard. The gray disks mark the metal cylinders and the red and blue disks mark the voids and centers between adjacent metal cylinders, respectively. The former correspond to the carbon atoms in graphene arranged on a honeycomb sublattice, and the latter are at the sites of the kagome sublattice. The honeycomb sublattice is terminated by armchair and zigzag edges along the short and the long sides, respectively and translationally invariant with respect to all sides.}
    \label{fig:schemebilliard}
\end{figure}

It was shown in Ref.~\onlinecite{raghu2008analogs} that under certain conditions photonic crystals with a triangular lattice geometry exhibit a linear Dirac dispersion relation. In Ref.~\onlinecite{dietz2015spectral} the spectral properties of a rectangular, superconducting microwave Dirac billiard were investigated experimentally. The microwave resonator~\cite{bittner2012extremal,dietz2013lifshitz} consisted of a metal lid and a basin containing $888$ cylinders which were milled out of another metal plate. The arrangement of the cylinders is illustrated schematically by gray disks in Fig.~\ref{fig:schemebilliard}.  
Below the microwave frequency $f_{max}=c/2h$ with $c$ denoting the velocity of light and $h$ the height of the cavity, the electric field excited inside the resonator is perpendicular to the top and bottom plates. Furthermore, it vanishes at the resonator and cylinder walls. Thus, the associated Helmholtz equation is two-dimensional and mathematically equivalent to the Schr\"odinger equation of the corresponding quantum billiard with Dirichlet boundary conditions.~\cite{stoeckmann1990quantum,sridhar1991experimental,graf1992distribution} This implies that the resonance frequencies yield the eigenvalues $k_n$ or, equivalently, the eigenfrequencies $f_n=\frac{k_nc}{2\pi}$ with $c$ denoting the velocity of light in vacuum, of a rectangular quantum billiard of corresponding shape containing circular scatterers at the positions of the cylinders subject to the same boundary conditions. The height of the Dirac billiards was $h = 3$ mm corresponding to $f_{max}=50$~GHz. A complete sequence of $\approx 4900$ resonance frequencies was determined below $f_{max}\approx 50$~GHz; see Ref.~\onlinecite{dietz2015spectral} for more details.

\begin{figure}[htb!]
\centering
    \includegraphics[width=0.9\columnwidth]{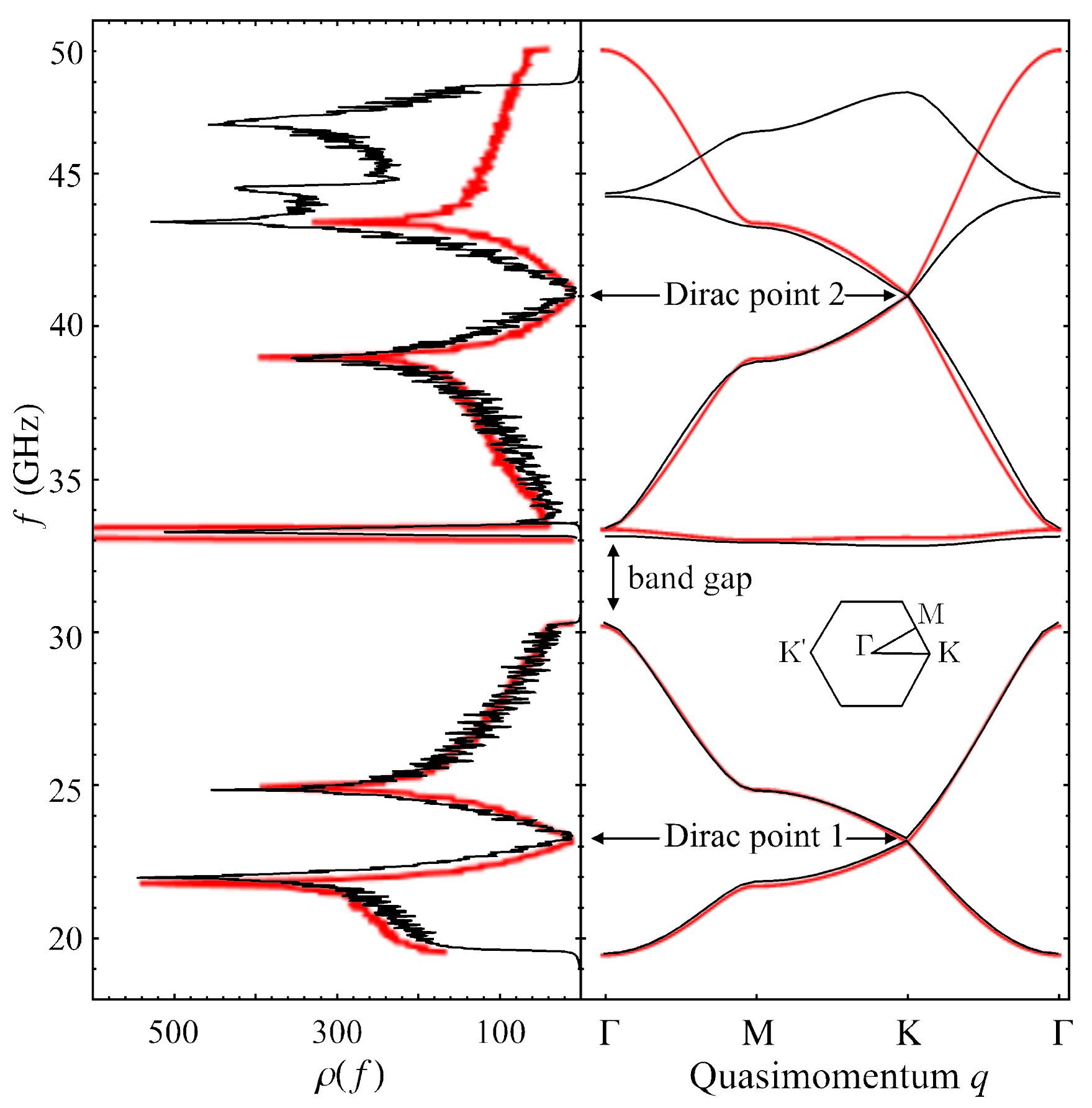}
	\caption{Left panel: Comparison of the experimental DOS (black) with the final TBM result (Model {\bf C} in Sec.~\ref{sec:results}) obtained for a honome lattice (red) combining honeycomb and kagome lattices. Right panel: Band structure of an infinite honeycomb (black) and infinite honome (red) lattices.  Here, we took into account up to 6th nearest-neighbor hoppings. The lattice constant of the honeycomb lattice was chosen equal to that of the void structure of the photonic crystal. The frequencies of the DPs, the band gaps and the Van Hove singularities and the other peaks of the experimental $\rho(f)$ agree well with the computed ones and the locations of the saddle points and the FB, respectively.} 
    \label{fig:dos}
\end{figure}

\begin{figure}[h]
\centering
	\includegraphics[width=0.9\columnwidth]{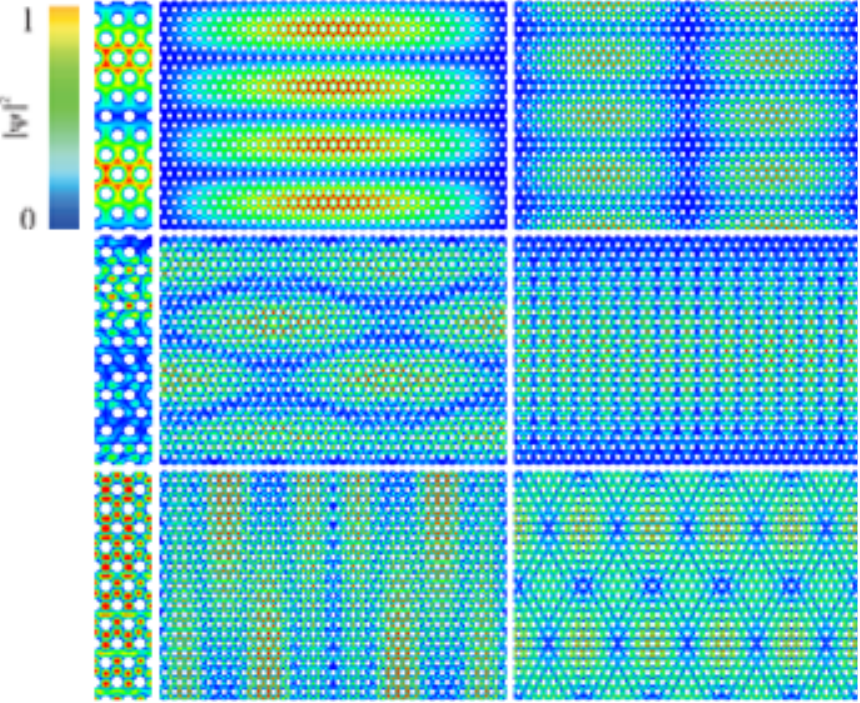}
	\caption{Electric-field intensity distributions in the microwave Dirac billiard corresponding to resonance frequencies in the bands framing the lower DP (top row), the FB (middle row) and the bands framing the upper DP (bottom row). In the region below the FB the electric-field intensity is maximal at the centers of the voids forming the honeycomb lattice and spreads into the neighboring ones, thus leading to an overlap of the electric-field (wave function) components localized there. In the region of the FB the electric-field intensity is localized between two adjacent cylinders corresponding to the locations of the kagome-sublattice sites in the honome-based TBM and in the region above the FB it can be maximal at both the honeycomb- and kagome-sublattice sites. This is better visible in the zoom into the left wave functions shown adjacent to it to the left. The color code is shown in the upper left corner.     
}
\label{fig:WFW}
\end{figure}

The left panel of Fig.~\ref{fig:dos} shows the density of states (DOS), $\rho (f)=\frac{\pi^2}{N}\sum_n\delta(f-f_n)$ with $N$ denoting the number of sites of the honeycomb lattice, i.e., of voids formed by the metal cylinders, respectively. 
It exhibits broad minima of low resonance density located around two DPs, which are framed by van Hove singularities.~\cite{vanhove1953the} These are separated by a narrow region of exceptionally high resonance density at the upper edge of the first band gap. Thus, the resonance spectra of microwave Dirac billiards exhibit DPs where they are governed by the relativistic Dirac equation.~\cite{bittner2010observation} The origin of these features lies in the honeycomb structure formed by the electric-field intensity patterns of the propagating modes which, below the FB exhibit maxima at the voids between three neighboring cylinders forming a triangular cell. The top row of Fig.~\ref{fig:WFW} shows spatial patterns of the electric-field intensity distribution for resonance frequencies from the bands framing the lower DP. They were computed by solving the two-dimensional Helmholtz equation~\cite{dietz2015spectral} with Dirichlet boundary conditions at the resonator and cylinder walls.~\footnote{Private communication with W. Ackermann, TEMF, TU-Darmstadt, Germany}
Hence, the voids marked by red disks in Fig.~\ref{fig:schemebilliard} correspond in graphene to the carbon atoms forming the honeycomb structure and the intensity of the electric field at the voids is related to the on-site excitations.~\cite{bittner2012extremal,dietz2019from} The void structure terminates with a zigzag edge along the longer edges of the rectangle and with an armchair edge along the shorter ones. The voids correspond to open resonators~\cite{Gaspard1989} and the electric-field intensity localized inside the cells may be considered as quasibound states which, depending on the resonance frequency, might partially overlap with neighboring ones as exhibited by the top left example in Fig.~\ref{fig:WFW}. 

Similarly, the frequencies of wave propagation as function of the two quasimomentum components exhibit a band structure, shown in the right panel of Fig.~\ref{fig:dos}, which resembles that of graphene around the DPs. The calculated band-structure function $f(\vec q)$ (black curves) is plotted along the path $\Gamma{\rm M}{\rm K}\Gamma$ inside the first Brillouin zone (BZ). Here, K and M denote the distances of the DPs located at the corners of the BZ and the saddle points from the $\Gamma$ point at the center of the BZ where the band terminates, respectively.~\cite{castro2009the} The van Hove singularities~\cite{vanhove1953the} correspond to the saddle points in the band structure. Generally, $\rho(f)$ exhibits maxima at frequencies corresponding to regions of low group velocity, $\vert\vec\nabla f(\vec q)\vert\simeq 0$ with $\vec q$ denoting the quasimomentum vector. Indeed, the narrow region of high resonance density is in the frequency range of the FB clearly visible in the band structure function. The positions of the experimental band gaps and DPs agree well with those in the calculated band structure. 

The qualitative behavior of the DOS as function of frequency is reminiscent of the DOS of a honome lattice. The red curves in the panels of Fig.~\ref{fig:dos} show the final results for the DOS and band structure obtained with a TBM for a honome lattice taking into account up to 6th nearest-neighbor hopping, where the lattice constant of the honeycomb sublattice had the same size as the lattice formed by the voids of the photonic crystal. Below the topmost van Hove singularity the calculated DOS agrees well with the experimental one. Asymmetries of the DOS in height and position of the van Hove singularities with respect to the associated DP and of the DPs including the bands framing them with respect to the narrow peak are attributed in general to longer-range hoppings as is outlined in Sec.~\ref{sec:model}. Note, that the TBM does not account for the topmost band which is beyond the validity of the honeycomb-based TBM used in Ref.~\onlinecite{dietz2015spectral} or the honome-based TBM used in the present article. Furthermore, the electric-field strength excited inside the resonator is no longer perpendicular already above $f\gtrsim 44$~GHz, implying that the Helmholtz equation becomes vectorial and thus the analogy with artificial graphene is lost, so that the TBM ceases to be applicable already below $f_{max}$.

\section{Tight-binding model for microwave photonic crystal}
\label{sec:TBM}

In Ref.~\onlinecite{dietz2015spectral} a TBM was used to describe the experimental DOS. In distinction to the first TBM description of graphene,~\cite{wallace1947the} where only nearest- and second-nearest-neighbour interactions of the $p_z$ orbitals were considered, also third-nearest-neighbour couplings and, in addition, overlaps between the wave functions centered at the associated atoms had to be included in order to attain good agreement between the experimental and computed DOS. The overlap parameters take into account the partial overlap of the quasibound states, i.e., of the electric field mode components localized at the voids formed by the metal cylinders with neighboring ones,~\cite{reich2002tight} indicating that it is nonvanishing between adjacent metal cylinders. Accordingly, the band-structure function $f(\vq)$ was obtained by solving the generalized eigenfrequency problem
\begin{gather}
    \label{eq:eigenfrequency}
    \mh_\text{TB}\vert\Psi_{\vq}(\vec r)\rangle=f(\vq)\mathcal{S}_{WO}\vert\Psi_{\vq}(\vec r) \rangle
\end{gather}
with the TBM Hamiltonian
\begin{gather}
    \label{eq:htb}
    \mathcal{H}_{TB} = \left({
    \begin{array}{cc}
        \gamma_0 + \gamma_2 h_2(\vq)\, & \gamma_1 h_1(\vq) + \gamma_3 h_3(\vq)\\
        \gamma_1 h_1(\vq) + \gamma_3 h_3(\vq)\, & \gamma_0 + \gamma_2 h_2(\vq)
    \end{array}}\right)
\end{gather}
and the wave function overlap matrix
\begin{gather}
    \label{eq:stb}
    \mathcal{S}_\text{WO} = \left({
    \begin{array}{cc}
        1 + s_2 h_2(\vq)\, & s_1 h_1(\vq) + s_3 h_3(\vq)\\
        s_1 h_1(\vq) + s_3 h_3(\vq)\, & 1 + s_2 h_2(\vq)
    \end{array}}\right)\, ,
\end{gather}
incorporating the n.n. coupling $\gamma_1$ and the 2nd and  3rd n.n. couplings $\gamma_2$ and $\gamma_3$ and the corresponding overlap parameters $s_1$, $s_2$ and $s_3$. The functions $h_n(\vq),\, n=1,2,3$ associated with the different couplings are given in Ref.~\onlinecite{reich2002tight}. The parameters were determined by fitting the DOS deduced from the band-structure function $f(\vec{q})$ to the experimental one. 

Since this TBM applies to infinitely extended hexagonal lattices and thus does not describe finite-size effects, in Ref.~\onlinecite{dietz2015spectral} as a further step a finite TBM for the bounded hexagonal lattice formed by the voids of the microwave photonic crystal was used to improve the agreement between the experimental DOS and the TBM description. In order to determine the associated DOS a $N\times N$ dimensional TBM matrix needed to be diagonalized, using the values for the parameters $\gamma_0,\, \gamma_n,\, n=1,2,3$ and $s_n,\, n=1,2,3$ obtained with the TBM Eq.~\eqref{eq:eigenfrequency} as initial values for the fitting procedure. Thereby, a good TBM description was achieved for the bands framing the lower DP. However, for the upper one the DOS deduced from the TBM needed to be fitted separately to the experimental one below and above the DP in order to achieve a good agreement between both curves. Furthermore, the honeycomb-lattice based TBM cannot describe the FB. Thus, it is not suitable to describe the band structure including both DPs.

\section{Improved Tight-Binding Model: the honome lattice}
\label{sec:model}

The fact, that the above TBM yields a good description of the DOS and fluctuation properties of the resonance frequencies in the bands framing the lower DP only after including wave function overlaps, already indicates that the electric-field strength, which has maximal intensity at the voids in that frequency range may extend into neighboring ones, implying that it is nonzero between adjacent metal cylinders. This is confirmed when inspecting the associated electric-field intensity distributions as illustrated in the top row of Fig.~\ref{fig:WFW} showing two typical examples. Furthermore, it is supported by the spatial patterns of the electric-field distributions corresponding to resonance frequencies in the FB and the bands framing the upper DP. In the region of the FB all electric-field distributions exhibit maxima between two adjacent cylinders. Around the upper DP the electric-field intensities are typically large at both the centers of the voids and between two adjacent cylinders. These observations and the features of the DOS, namely  the occurrence of a narrow region of particularly high resonance density between two DPs framed by van Hove singularities led us to the idea that the microwave Dirac billiard, actually, might provide an experimental realization of a honome lattice. In this picture, the non-vanishing electric field intensity at the void centers and between two adjacent metal cylinders, marked by red and blue dots, respectively, in Fig.~\ref{fig:WFW} correspond to atoms of the honeycomb and kagome sublattice, respectively. Nevertheless, in accordance with the spatial patterns of the electric-field intensities, illustrated in Fig.~\ref{fig:WFW}, the lower DP including the bands framing it is well described by the TBM for a finite-size honeycomb sublattice, implying that the coupling between the sublattices is weak. Note, that the features of the spatial patterns of the electric-field intensity in the regions of the FB and above the FB cannot be accounted for by a honeycomb-base comprising more basis functions at the honeycomb lattice sites~\cite{Marzari1997,Marzari2012,Jung2013} using the results of Ref.~\onlinecite{Gaspard1989} on three-disk scattering, as was done, e.g., to describe those in the microwave-billiard realization of fullerene.~\cite{Dietz2015,dietz2019from} The wave function patterns can be understood as the result of interferences of the waves of the elastic multiscattering process off the triangular arrangement of cylinders. Their characteristics are dictated by the symmetry properties of the graphene billiard, shown schematically in~\ref{fig:schemebilliard}. 
\begin{figure}[t!]
    \includegraphics[width=1\columnwidth]{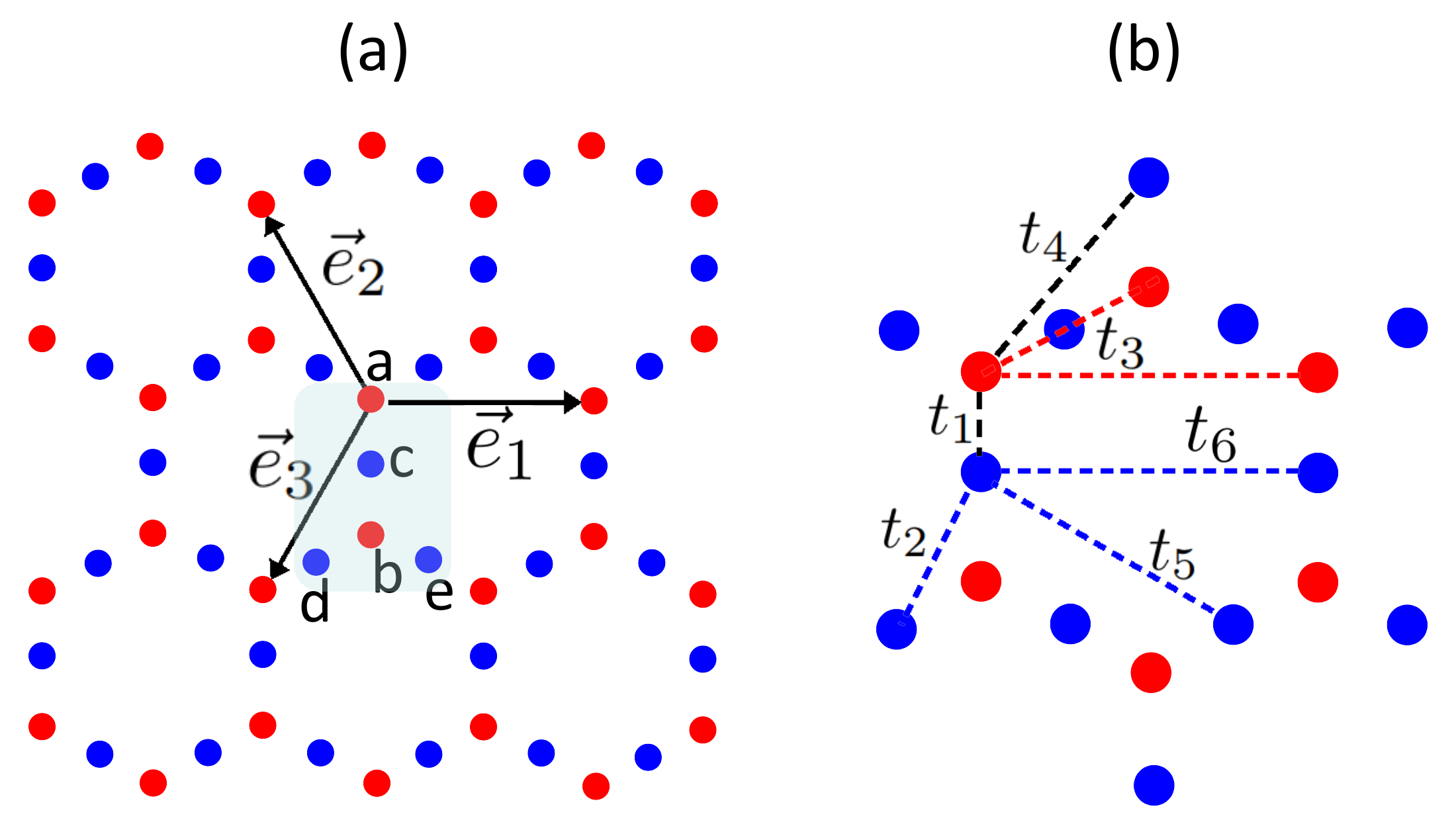}
	\caption{The honome lattice, consisting of two sublattices -- honeycomb (red sites) and kagome (blue sites). (a) Structure of the honome lattice. The shaded area marks the unit cell. $\ve_1$ and $\ve_2$ are the basis vectors generating the lattice and $\ve_3$ equals their negated sum. (b) The hoppings taken into account in the present article. The hopping parameters $t_j$ are ordered according to their distances. Hoppings between the sites of the honeycomb sublattice are plotted as red dashed lines, between the sites of the kagome sublattices as blue dashed line and between the sites of the two sublattices as black dashed lines.}
    \label{fig:honome}
\end{figure}

We will demonstrate in the following that a modified TBM defined on a honome lattice indeed is capable to reproduce the overall structure of the wave function patterns for the bands framing the two DPs and above all, in the FB region. It is constructed by placing kagome sites at the centers of the honeycomb n.n. bonds as demonstrated in Fig.~\ref{fig:honome}.~\footnote{Note, that the kagome lattice is the line graph of the honeycomb lattice} We refer to the honeycomb sites of the honome lattice as honeycomb sublattice, and to the kagome sites as the kagome sublattice. Given the unit cell translation vectors $\ve_1,\ve_2$ (see Fig.~\ref{fig:honome}(a))
\begin{gather}
    \ve_1 = \left(1, 0\right),\quad \ve_2 = \left(-\frac{1}{2}, \frac{\sqrt{3}}{2}\right),\\
    \ve_3 = -\ve_1 - \ve_2 = \left(-\frac{1}{2}, -\frac{\sqrt{3}}{2}\right),
\end{gather} 
the unit cells are labelled by two integers $m, n\in\mathbb{Z}$, which give the position $\vec{R}_{mn}= m \ve_1 + n \ve_2$ of that cell. The wave function amplitudes are labeled by the indices $n,m$ of the unit cell they belong to, while the amplitudes within the unit cells are labeled by $a,\, b,\, c,\, d$ and $e$, respectively,~\cite{maimaiti2017compact,maimaiti2019universal}
\begin{gather}
    \vpsi_{mn} = \left(a_{mn}, b_{mn}, c_{mn}, d_{mn}, e_{mn}\right)^T.
\end{gather} 
Our convention is that $a_{mn}, b_{mn}$ correspond to the wave function amplitudes on the honeycomb sites and $c_{mn}, d_{mn}, e_{mn}$ to those on the kagome sites (see Fig.~\ref{fig:honome}). The honeycomb-based TBM~\cite{dietz2015spectral} provided an adequate description for the lower part of the spectrum with up to 3rd n.n. hoppings on the honeycomb lattice which corresponds to 8th n.n. hoppings on the honome lattice. Since it turned out that the description of the experimental DOS is already good for up to 6th n.n. hoppings, we restrict ourselves to this hopping range as illustrated in Fig.~\ref{fig:honome} (b). It corresponds to up to 2nd n.n. hoppings in the honeycomb sublattice (red dashed lines), up to 3rd n.n. in the kagome sublattics (blue dashed lines) and up to 2nd n.n. hoppings between the sublattices (black dashed lines). We adopt the conventions of Refs.~\onlinecite{maimaiti2017compact,maimaiti2019universal,maimaiti2020thesis} and describe the hopping in terms of matrices that give the hopping inside the unit cell or between different unit cells. These matrices read in our case
\begin{equation}
    \begin{aligned}
        H_0 = \begin{pmatrix}
            \epsilon & t_3 & t_1 & t_4 & t_4\\
            t_3 & \epsilon & t_1 & t_1 & t_1\\
            t_1 & t_1 & 0 & t_2 & t_2\\
            t_4 & t_1 & t_2 & 0 & t_2\\
            t_4 & t_1 & t_2 & t_2 & 0
        \end{pmatrix},\ \ 
        & H_1 = \begin{pmatrix}
            t_6 & 0 & 0 & 0 & 0\\
            0 & t_6 & t_4 & t_4 & 0\\
            0 & 0 & t_6 & t_5 & 0\\
            0 & 0 & 0 & t_6 & 0\\
            0 & t_4 & t_5 & t_2 & t_6
        \end{pmatrix}\\
        H_2 = \begin{pmatrix}
            t_6 & t_3 & t_4 & t_4 & t_1\\
            0 & t_6 & 0 & 0 & t_4\\
            0 & t_4 & t_6 & t_5 & t_2\\
            0 & 0 & 0 & t_6 & t_5\\
            0 & 0 & 0 & 0 & t_6
        \end{pmatrix},\ \
        & H_3 = \begin{pmatrix}
            t_6 & 0 & 0 & 0 & 0\\
            t_3 & t_6 & t_4 & 0 & 0\\
            t_4 & 0 & t_6 & 0 & 0\\
            t_1 & t_4 & t_2 & t_6 & t_5\\
            t_4 & 0 & t_5 & 0 & t_6
        \end{pmatrix}
    \end{aligned}
    \label{eq:hop-mats}
\end{equation}
\noindent where $t_j$ is the $j$th n.n. hopping parameter and $\epsilon$ is the onsite energy on the honeycomb sublattice. Here, $H_0$ is the intra unit cell hopping matrix and $H_{i=1,2,3}$ describes hoppings between n.n. unit cells along the lattice vectors $\ve_i$. From the fact, that the honeycomb-lattice based TBM yields a good description below the FB and inspection of the spatial electric-field intensity patterns in that frequency range, showing that they are larger at the centers of the voids, i.e., on the honeycomb sites, than between adjacent metal cylinders, i.e., on the kagome sites we may conclude that the onsite energies differ for the two sublattices. We use the freedom to fix the zero of energy to shift the onsite energy on the kagome sublattice to zero leaving only the honeycomb sublattice onsite energy $\epsilon$ as parameter. We also set the n.n. hopping parameter $t_1=1$, thereby fixing the overall scale of energy. The eigenfrequency problem for the honome-based TBM reads
\begin{gather}
    \label{eq:eig-prob}
    H_0 \vpsi_{mn} + H_1 \vpsi_{m+1,n} + H_1^\dagger \vpsi_{m-1,n} + H_2 \vpsi_{m,n+1} + \\
    H_2^\dagger \vpsi_{m,n-1} + H_3 \vpsi_{m+1,n+1} + H_3^\dagger \vpsi_{m-1,n-1} = E \vpsi_{mn}.\notag
\end{gather}
Due to the translational invariance, the eigenvectors are plane waves $\vpsi_{mn} = \vec{\phi}_{\vk} e^{-i \vk \cdot \vec{R}_{mn}}$ with $\vk = (k_x, k_y)$. Plugging this ansatz into Eq.~\eqref{eq:eig-prob} the phase factor $e^{-i \vk \cdot \vec{R}_{mn}}$ drops out yielding the eigenfrequency equation 
\begin{gather}
    \label{eq:eig-prob1}
	\mh_k\vec{\phi}_{\vk}=E(\vk)\vec{\phi}_{\vk}
\end{gather}
with the $k$-dependent $5\times 5$ Hamiltonian
\begin{gather}
    \label{eq:hk}
    \mh_k = H_0 + \sum_{j=1}^3\left(H_j e^{-i k_j} + H_j^\dagger e^{i k_j}\right),
\end{gather}
where $k_i = \vk\cdot\ve_i$. The eigenfrequencies $E(\vk)$ of $\mh_k$ provide the full band structure from which we compute the DOS~\cite{Hobson1953} 
\begin{gather}
	\rho(\omega)=\frac{1}{\mathcal{A}_{\rm BZ}}\iint_{\rm BZ}\delta[\omega-E(\vk)]{\rm d}k_x{\rm d}k_y,
\end{gather}
with $\mathcal{A}_{\rm BZ}$ denoting the area of the BZ.
Our aim is to find the set of hopping parameters $t_j$ and onsite energy $\epsilon$ for which the TBM DOS best fits the experimental DOS. 

Before proceeding further, it is useful to discuss the basic properties of the simplest TBM taking into account only n.n. hoppings on the honome lattice, $t_1\ne 0$, $t_j=0$ for $j\geq 2$. In that case, the TBM has a perfect FB at $E=0$ and its DOS exhibits a sharp peak of vanishingly small width at $E=0$.   This is a consequence of the chiral symmetry, namely, the honome lattice is bipartite and the n.n. model has $5$ bands.~\cite{ramachandran2017chiral} As we will see below this FB is the origin of the sharp peak of the DOS in Fig.~\ref{fig:dos} implying that significant remnants of the FB survive, even in presence of a perturbation which breaks the lattice symmetry that induces the appearance of the FB. 

\section{Methods}
\label{sec:methods}

In order to find the set of hopping parameters for which the honome TBM yields the best description of the resonance frequencies and electric field distributions of the microwave Dirac billiard, we compute the DOS of the honome TBM for a given set of hoppings $t_j$ and onsite energy $\epsilon$ and compare it with the experimental DOS to identify the set for which agreement was best. For this, we fit the positions and the weights of characteristic features of the TBM DOS -- the van Hove singularities, the FB and the DPs -- to their experimental counterparts. The model~\eqref{eq:hk} contains $6$ free parameters, $t_j,\, j=2,\dots , 6$ and $\epsilon$. In order to find their values we successively fit the TBM with increasing number of hopping parameters to the experimental DOS, that is, we start with a TBM Hamiltonian containing only $t_2,\; t_3$ and $\epsilon$ as fit parameters and increase the hopping range until satisfactory results are achieved. For convenience we use the Fourier space Hamiltonian~\eqref{eq:hk} for all the computations. Alternatively one could use the real-space analog of~\eqref{eq:eig-prob} for a finite-size TBM. However, the associated fitting procedure is more time-consuming, because it requires the diagonalization of a Hamiltonian matrix of which the dimension is given by the number of sites.~\cite{dietz2015spectral} To check consistency of the two fitting procedures we inserted the resulting parameters into the real-space Hamiltonian reproducing the associated DOS.

\subsection{Rescaling of the spectra}
\label{sec:rescaling}

\begin{figure}[htb!]
    \centering{}
    \includegraphics[width=0.9\columnwidth]{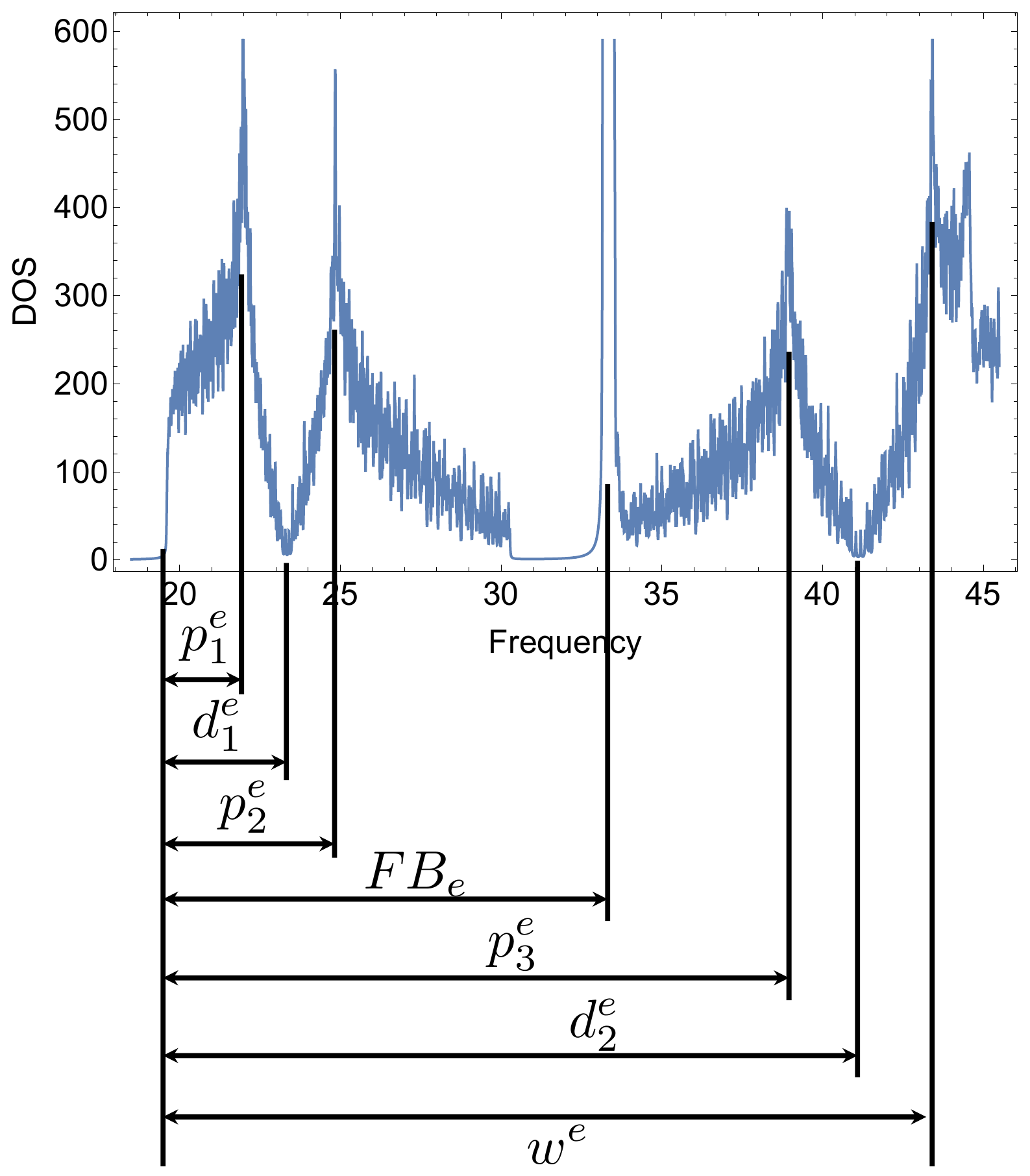}
	\caption{The positions $p^e_i,\, i=1,\dots , 3$ of the peaks, of the FB $FB^e$ and the DPs $d^e_i,\, i=1,2$ with respect to the lower band edge for the experimental DOS.}
    \label{fig:exp_dos_bandstr}
\end{figure}

The frequency range of the experimental DOS and the energy scale of the TBM calculations differ. Accordingly, we need to eliminate the scale. We define the relative position (RP) of the van Hove singularities $p_i^e,\, i=1,\dots , 3$ as 
\begin{gather}
    \label{eq:vhs-rel-pos}
    p_i^e = \frac{\omega_i - \omega_0}{w^e}.
\end{gather} 
Here, $\omega_i,\, i=1,\dots , 3$ denote the resonance frequencies of the von Hove singularities, $\omega_0$ that of the band edge below the lower DP and $w^e$ the frequency range from $\omega_0$ to the van Hove singularity above the upper DP in the experimental DOS as depicted in Fig.~\ref{fig:exp_dos_bandstr}. Similarly, $d_i^e,\, i=1,2$ are the positions of the Dirac points and $\text{FB}^e$ that of the FB with respect to $\omega_0$, rescaled with $\omega^e$. The values of $p_i^e$, $d_i^e$ and $\text{FB}^e$ are summarized in Table~\ref{tab:exp-rel-pos}. To compute the width of the spectrum $w^e$ we took the value of the lower edge provided in Ref.~\onlinecite{dietz2015spectral},  $\omega_0=19.64$~GHz. For the upper edge of the spectrum we used the position of the $4$th van Hove singularity in the experimental DOS, $p_4^e = 43.421$~GHz. This choice reflects the fact that beyond this value the electric-field excited inside the resonator becomes three-dimensional, that is, the associated Helmholtz equation is no longer scalar, so that the approximate description in terms of a TBM~\cite{dietz2015spectral} fails. Furthermore, the honome-based TBM exhibits only five bands and thus, as expected,  does not account for the occurrence of the new band above the topmost van Hove singularity; see Fig.~\ref{fig:dos}, which, actually, would go beyond the purpose of this article. Accordingly, the width of the spectrum is given as
\begin{gather}
	w^e = (43.421 - 19.64)\text{GHz} = 23.781\text{GHz}.
\end{gather}

\begin{table}[htb!]
    \footnotesize
    \centering
    \begin{tabular}{|c|c| c| c| c| c| c|}
    \hline
       VHS  & $p_1^e$ & $p_2^e$ & $p_3^e$ & $FB_e$ & $d_1^e$ & $d_2^e$ \\ 
       \hline
	    RP (GHz)& 0.098524  & 0.218872 & 0.809259 & 0.573672 & 0.15586 & 0.902927 \\
        \hline
    \end{tabular}
    \caption{Relative positions (RP) of van Hove singularities, the FB and the DPs extracted from the experimental resonance frequencies in units of GHz.}
    \label{tab:exp-rel-pos}
\end{table} 

The relative positions of the van Hove singularities $p_i$, the FB and the DPs $d_i$ are defined similarly for the honome TBM. Here, we assume that the peaks of the van Hove singularities are located at the $M$ points, the DPs are located at $K$ points, and the FB and the band edges are located at the $\Gamma$ points associated with the two DPs. This is an approximation, justified by the assumption of weak interlattice coupling, except for the case of uncoupled honeycomb and kagome sublattices. The width of the spectrum of the TBM is defined as
	\begin{gather}
	    \label{eq:span_of_spec}
	    w = \max\left(\{p_i\}\right) - \gamma,
	\end{gather}
where $\gamma$ is the lowest band edge, which we take as the smallest eigenfrequency of $\mh_k$ at the $\Gamma$ point ($k_x=k_y=0$). The $p_i$ are the positions of the five peaks comprising the van Hove singularities and the FB in Fig.~\ref{fig:exp_dos_bandstr}. 

\subsection{Quantifying the deviations between the experimental and the TBM DOS}

Once we determined the two sets of relative positions of the van Hove singularities, the FB and the DPs, we compute the distances between their numerical and experimental values
\begin{align}
    \Delta d_i & = \vert d_i - d_i^e\vert,\\ 
    \label{eq:diff_with_exp}
    \Delta p_i & = \vert p_i - p_i^e\vert,\\ 
    \Delta_\text{FB} & = \vert E_\text{FB} - E_\text{FB}^e \vert\;.
\end{align}
The discrepancy between the experimental and tight-binding DOS on the honome lattice is quantified by the magnitude of these quantities. We used two metrics to identify the set of parameters which provides the best fit to the experimental DOS,
\begin{gather}
    \label{eq:standard_deviation}
    \Delta_\infty = \max(\Delta d_i, \Delta p_i, \Delta_\text{FB}),\\
    \Delta_2 = \sqrt{\sum_i\Delta d_i^2+ \sum_i\Delta p_i^2 + \Delta_\text{FB}^2}.
\end{gather}
The best fit is obtained by minimizing either $\Delta_\infty$ or $\Delta_2$. Both metrics provided qualitatively similar results. Therefore, in what follows, we restrict the results to the $\Delta=\Delta_\infty$ measure.

\subsection{The reverse Monte-Carlo algorithm}
\label{sec:algorithm}

We used two versions of the reverse Monte-Carlo (rMC)~\cite{mosegaard2002monte,dunn2011exploring} to identify the best set of $t_j$ and $\epsilon$ that minimize $\Delta$, the plain (plain rMC) and importance (importance rMC) samplings. The general best-fit procedure may be summarized as follows: 
\begin{enumerate}
    \item Set the hopping range/number of hoppings $t_j$.
    \item Pick initial hopping values, either random, or reuse the results from previous runs of the rMC where the size of the hopping parameters $t_j$ are assumed to decrease with increasing hopping distance, i.e., with increasing index $j$. 
    \item Compute eigenfrequencies of $\mh_k$ at the $\Gamma$, $M$ and $K$ points and compute the metric $\Delta$~\eqref{eq:standard_deviation}.
    \item \textit{Plain sampling}: generate a random new set of hoppings $t_j$ and onsite energy $\epsilon$.\\
          \textit{Importance sampling:} generate a random shift of a randomly chosen hopping $t_j$ or onsite energy $\epsilon$.
    \item Recompute the eigenfrequencies of $\mh_k$ at the $\Gamma$, $M$ and $K$ points and recompute the metric $\Delta$~\eqref{eq:standard_deviation}.
    \item Compare the updated value of the metric $\Delta$ to the value computed on the previous iteration. If the new $\Delta$ is smaller accept the new set of hoppings and onsite energies, otherwise discard them.
    \item Go to step 4 until a desired discrepancy $\Delta$ is achieved.
\end{enumerate}
The second assumption is made in light of the experimental situation, where the hoppings correspond to the overlaps between the electric-field (wave-function) components localized at the sites of the honeycomb and kagome sublattices, respectively, which decrease with increasing distance.

\section{Reverse Monte-Carlo results} 
\label{sec:results}

In the present section we report on the rMC fitting results of the experimental DOS by the DOS of the honome TBM: We started the fitting with the model with up to 3rd n.n. hoppings (model {\bf A}) and then gradually increased the hopping range - up to 5th n.n. hoppings (model {\bf B}), and 6th hoppings (model {\bf C}) to improve the agreement with the experimental data. Model {\bf A} has a perfect FB for any choice of the parameters. Indeed, as can be seen in Fig.~\ref{fig:honome}(b) $t_1$ couples n.n. sites of the kagome and the honeycomb sublattices, while $t_2$ and $t_3$ couple n.n. sites within the honeycomb and kagome sublattices, respectively. Consequently, the usual compact localized eigenstates of the n.n. TBM for a kagome lattice, for which the wave-function amplitudes are nonzero only on six sites forming a hexagon,~\cite{bergman2008band} are also eigenstates of  model {\bf A}. Due to the translational invariance of both models an extensive set of eigenstates can be generated by shifting the hexagon with nonzero wave-function components along the lattice, thus implying the existence of a FB. Hence, for model {\bf A} the FB is associated with a sharp peak of vanishingly small width that is present for any choice of parameters, while the experimental peak is broadened. This implicates the necessity to increase the hopping range in order to broaden the FB and to attain a good description of the experimental DOS. The model with up to 4th n.n. hoppings did not provide statisfactory results with the reverse Monte-Carlo fitting, while model {\bf B} with up to 5th n.n. hoppings yielded a better qualitative agreement with the experimental DOS. To obtain an even better agreement between the TBM and the experimental DOS, in particular at the lower edge of the band gap below the FB, we considered a further hopping range, up to $t_6$ in model {\bf C}. Note, that the latter model is the direct extension of the honeycomb-based TBM introduced in Ref.~\onlinecite{dietz2015spectral}. It is important to note here, that there are multiple sets of hopping parameters $t_j$ and onsite energy $\epsilon$ yielding the best fit for model {\bf A}. However, in models {\bf B} and {\bf C}, the set of parameters resulting from the best fit to the experimental DOS is robust up to numerical errors. This indicates that at least 5th to 6th n.n. hoppings are required to attain a reliable description of the experimental DOS. 

Generally, in order to determine the best fit to the experimental data we first applied the plain rMC. Its results were subsequently used as an input to the importance rMC for further optimization of the fit. Both the plain rMC and the importance rMC yielded good agreement with the experimental DOS. Accordingly, we present below only results from the importance sampling procedure. 

The set of optimal parameters $\{\epsilon, t_2,\dots, t_l \}$, where $l=3, 5, 6$ for models {\bf A}, {\bf B}, {\bf C} respectively, yields a rescaled DOS $\rho(\omega;\epsilon, t_1=1, t_2,\dots, t_l)$. For comparison with the experiment this DOS needs to be scaled back (see Sec.~\ref{sec:rescaling}):
\begin{align}
    \tilde{\rho} (\omega;\epsilon, t_1=1, \dots, t_l) &= r \rho (f;\epsilon, t_1=1, \dots, t_l) + d, \\
    r &= \frac{w_e}{w}, \quad d = 19.64 - \gamma .
\end{align}

\subsection{Model A}

\begin{figure}[htb!]
    \centering
    \includegraphics[width=0.58\columnwidth]{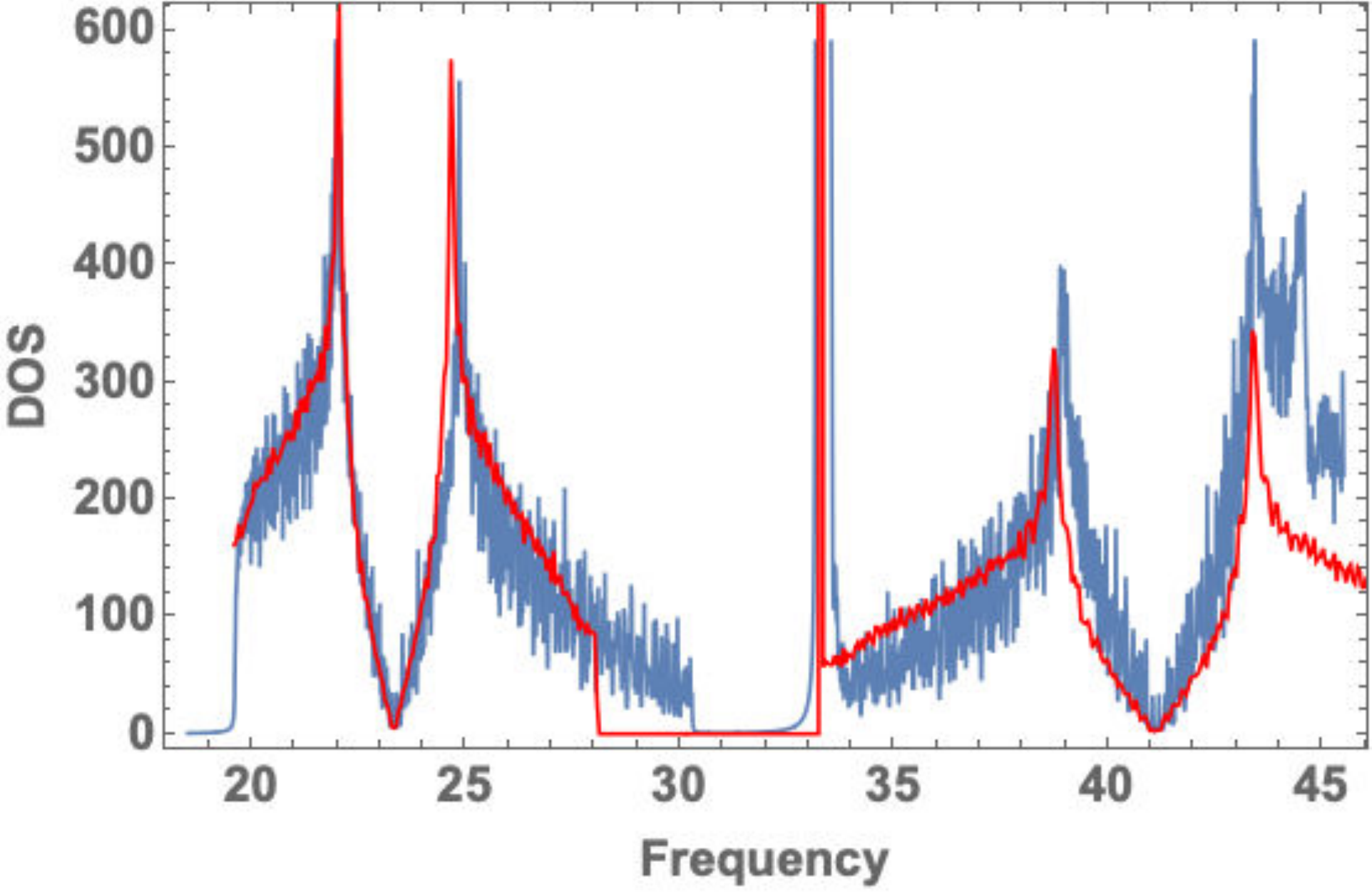}
    \includegraphics[width=0.38\columnwidth]{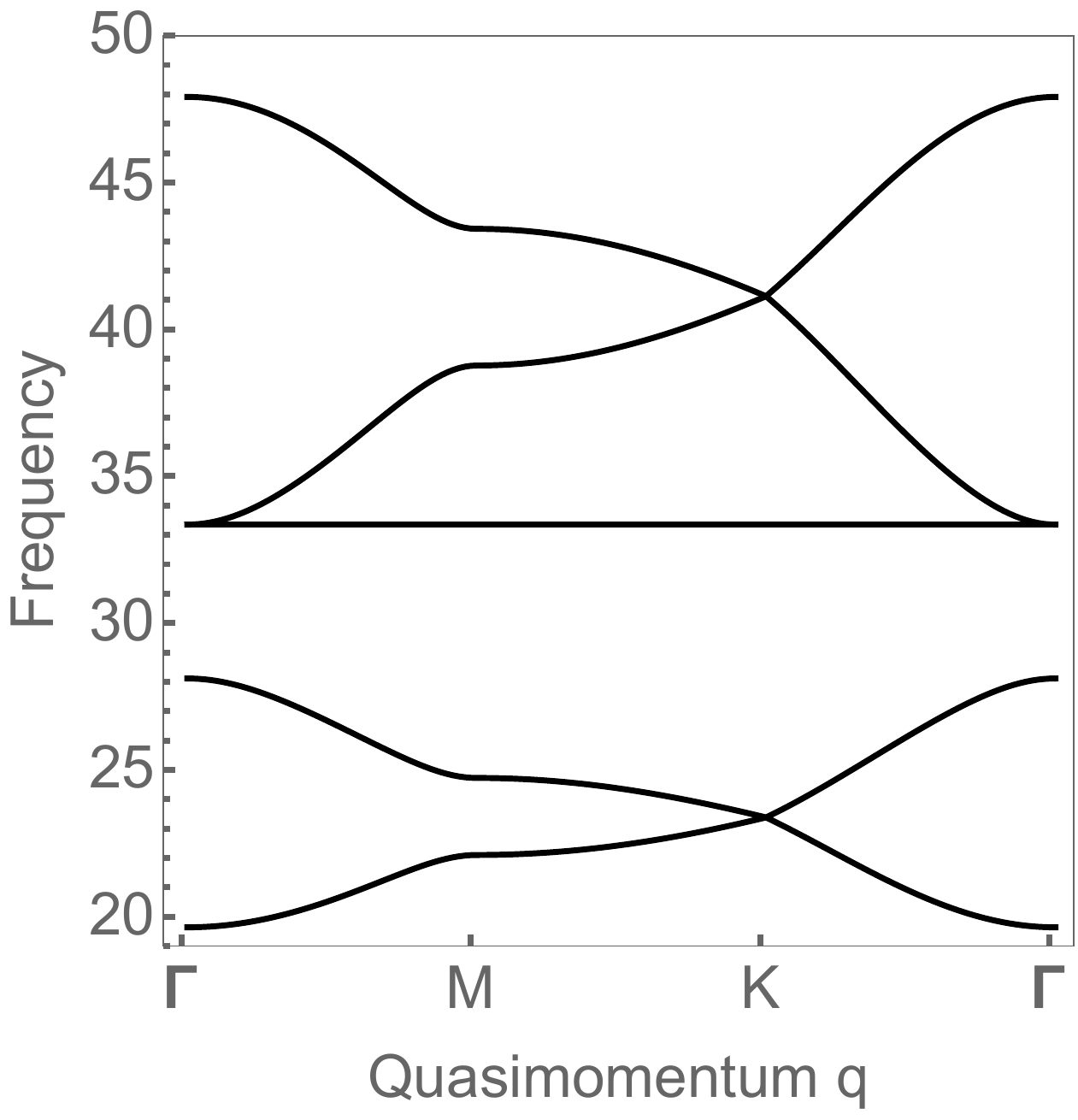}
	\caption{Best-fit results for model {\bf A} introduced in the main text obtained from the importance rMC by minimizing the metric $\Delta_\infty$. Left panel: The DOS of the optimized model {\bf A} (red) and the experimental DOS (blue). The FB peak is too sharp, and the upper edge of the second band framing the lower DP does not agree well with the experimental one. Right panel: The band structure of model {\bf A} along the path $\Gamma-M-K-\Gamma$ in the BZ.}
    \label{fig:imp_samp_fitting_3hoppings-1}
\end{figure}

In model {\bf A} only $t_1=1, t_2$ and $t_3$ are non-zero in Eq.~\eqref{eq:hop-mats}. In this case the eigenfrequencies of the Hamiltonian~\eqref{eq:hk} are known analytically at the $\Gamma$, $M$ and $K$ points. 
The fitting of the DOS to experimental data with the importance rMC is shown in  Fig.~\ref{fig:imp_samp_fitting_3hoppings-1}. The values of the optimal parameters are $t_1 = 1$ and $(\epsilon, t_2, t_3) = (-4.88694, 0.812533, -0.43105)$. The overall agreement is reasonable, however the FB peak is too sharp, because of the perfect FB, as was anticipated above. Therefore we conclude that the range of hoppings has to be extended to broaden the peak associated with the FB and achieve a better description of the experimental DOS. 

\subsection{Model B}

\begin{figure}[htb!]
    \centering
    \includegraphics[width=0.58\columnwidth]{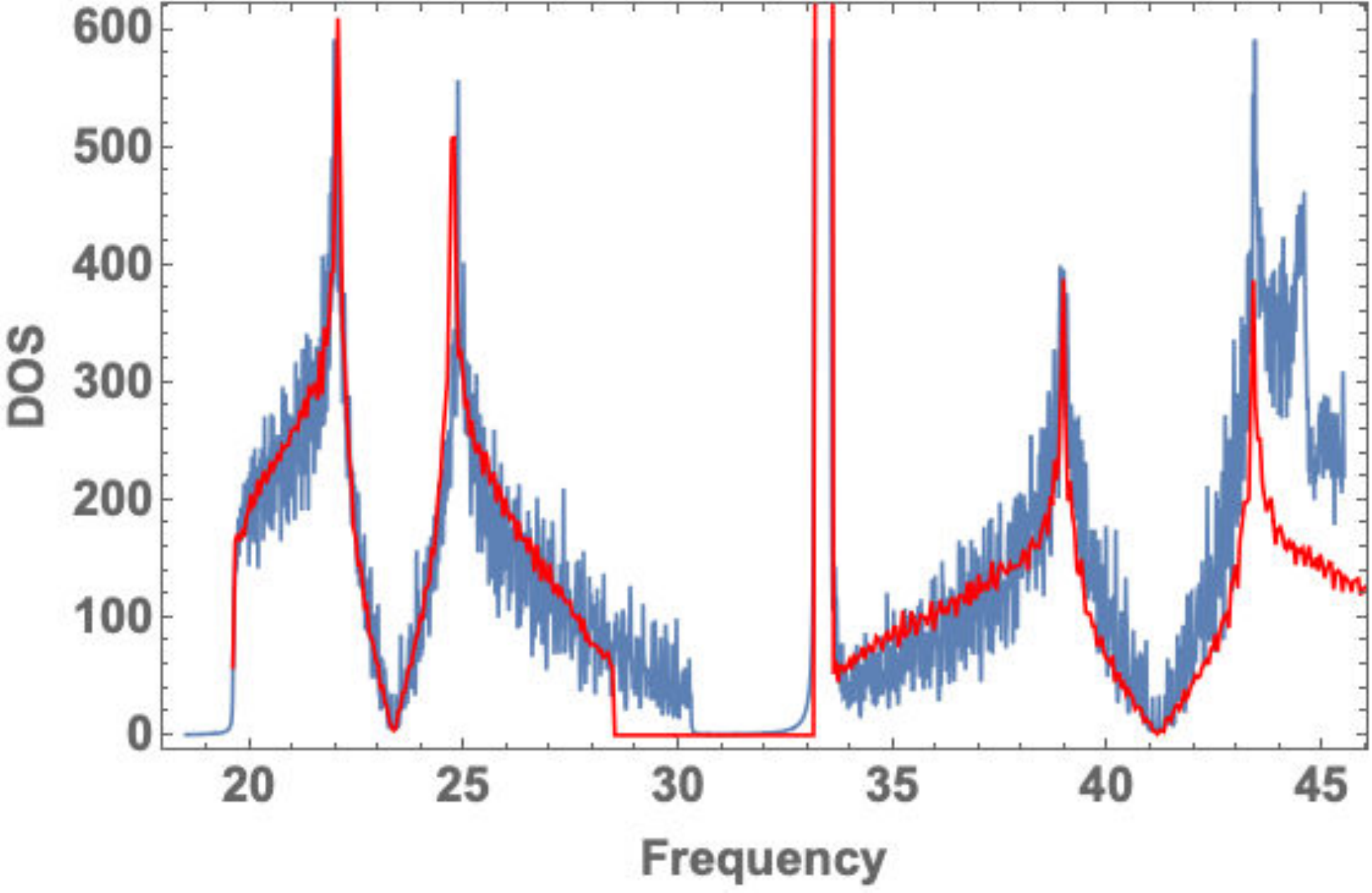}
    \includegraphics[width=0.38\columnwidth]{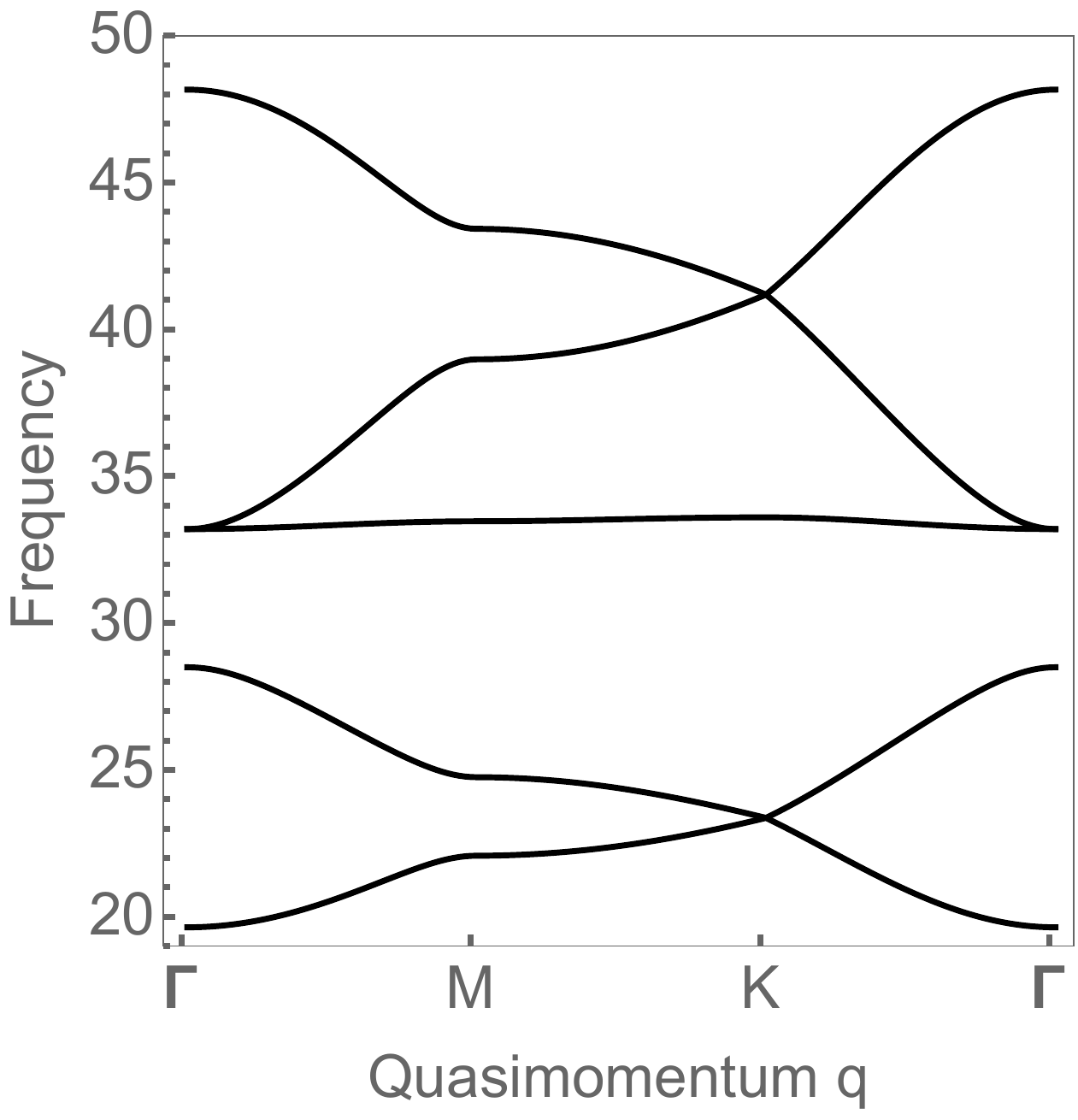}
	\caption{Best-fit results for model {\bf B} introduced in the main text obtained from the importance rMC by minimizing the metric $\Delta_\infty$. Left panel:  The DOS of the optimized model {\bf B} (red) and the experimental DOS (blue). The FB peak has broadened as compared to model {\bf A}, but the upper edge of the second band framing the lower DP still does not agree well with the experimental one. Right panel: The band structure of model {\bf B} along the path $\Gamma-M-K-\Gamma$ in the BZ.} 
    \label{fig:imp_samp_fitting_5hoppings-1}
\end{figure}

In model {\bf B} the hopping parameters $t_i, i=1,\dots , 5$ are non-zero in Eq.~\eqref{eq:hop-mats}. In this case the eigenfrequencies of the Hamiltonian~\eqref{eq:hk} can still be computed analytically at the $\Gamma$ and $M$ points but only numerically at the $K$ points. The result of the importance rMC fit is shown in Fig.~\ref{fig:imp_samp_fitting_5hoppings-1}. The values of the optimal parameters are $t_1=1$ and $(\epsilon, t_2, t_3, t_4, t_5)=(-5.02261, 0.852182, -0.486464, -0.04, 0.0256164)$. We observe a better agreement with the experimental data, including the broadening of the FB peak as compared with model {\bf A}. However this is still not true for the upper band edge of the second band framing the lower DP, suggesting that a larger hopping range needs to be considered.

\subsection{Model C}

\begin{figure}[htb!]
    \centering
    \includegraphics[width=0.58\columnwidth]{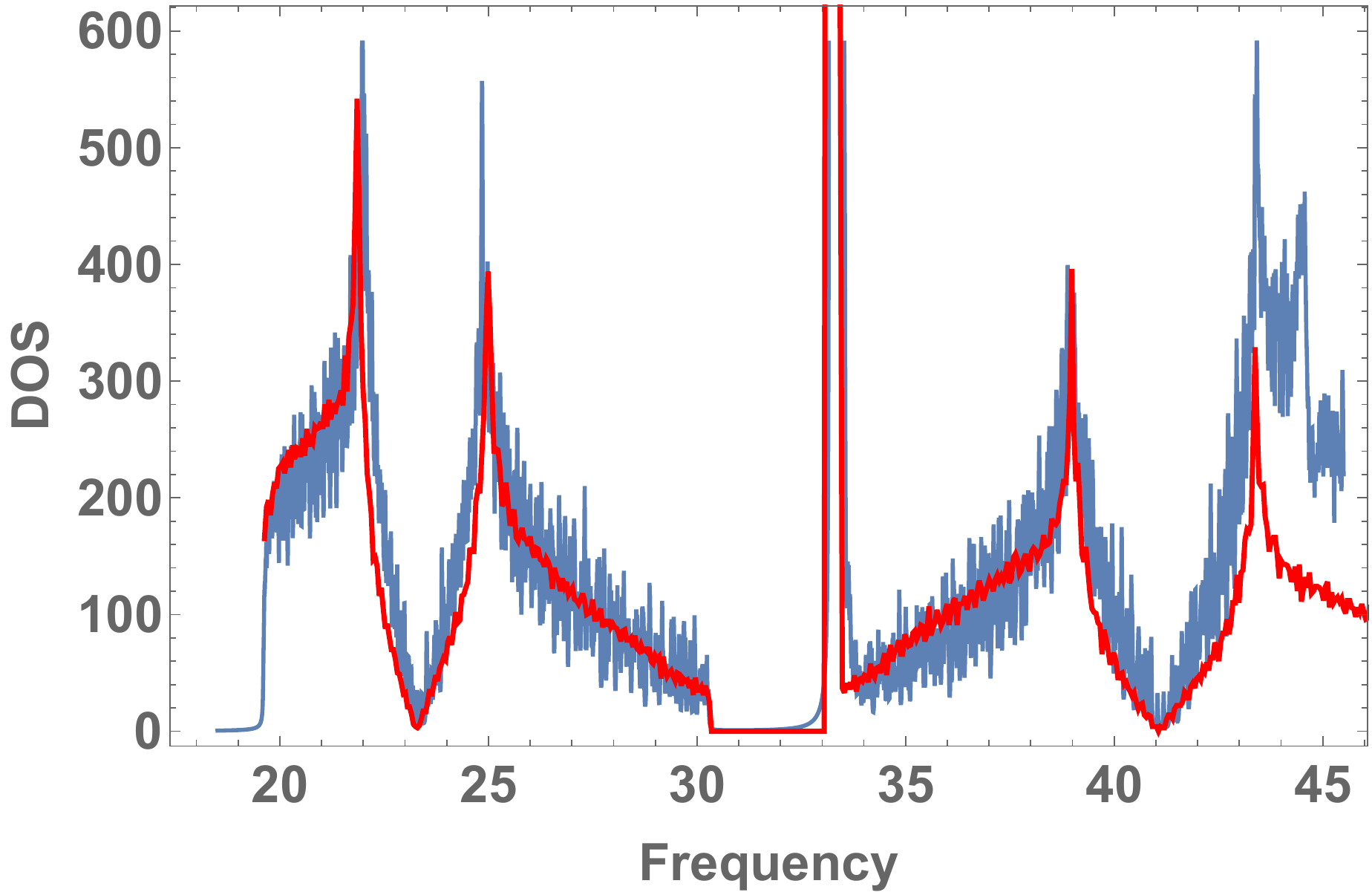}
    \includegraphics[width=0.38\columnwidth]{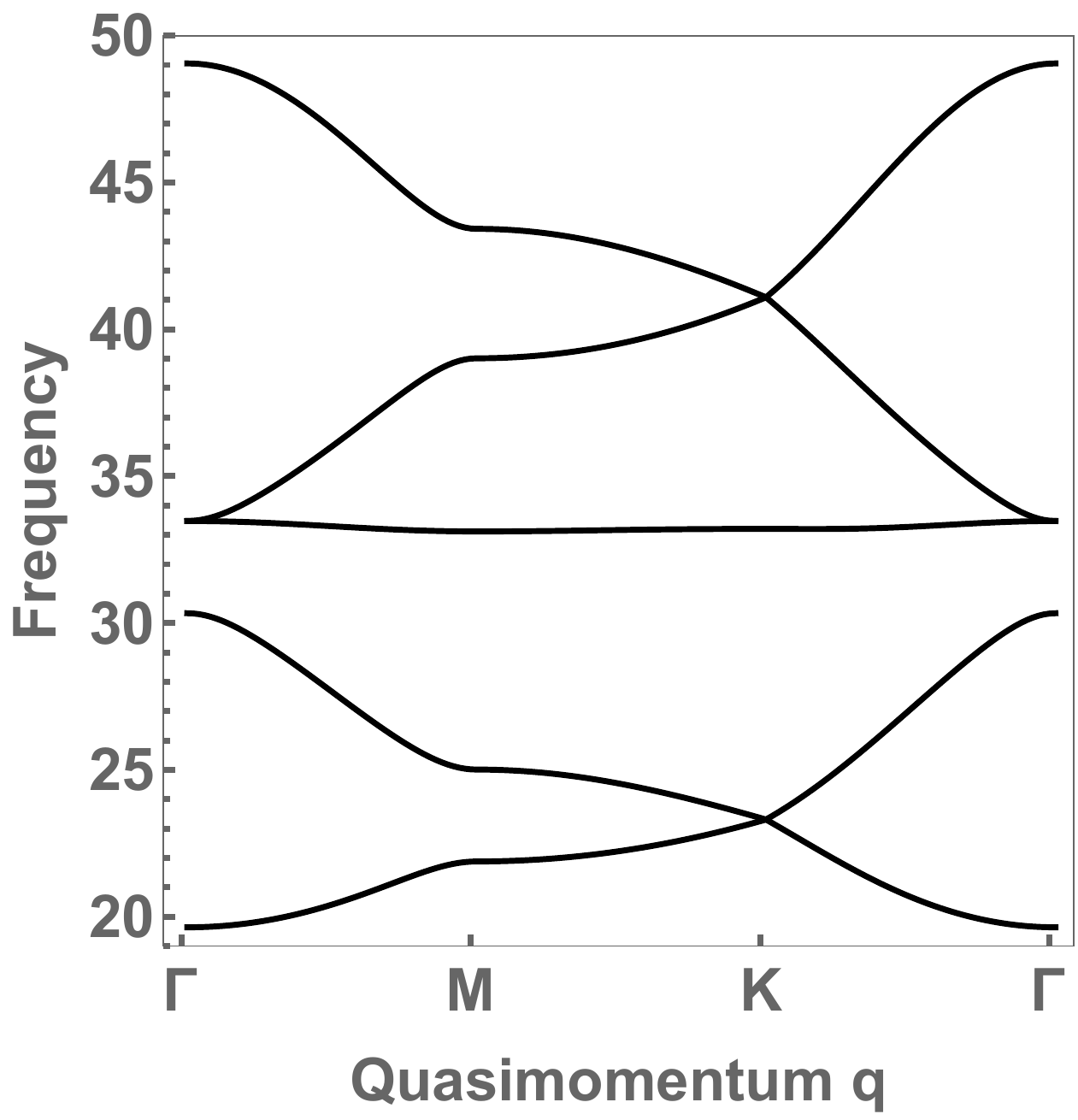}
	\caption{Best-fit results for model {\bf C} introduced in the main text obtained from the importance rMC by minimizing the metric $\Delta_\infty$. Left panel: The DOS of the optimized model {\bf C} (red) and the experimental DOS (blue). The FB peak has broadened as compared to model {\bf A} and the upper edge of the second band framing the lower DP agrees well with the experimental DOS. Right panel: The band structure of model {\bf C} along the path $\Gamma-M-K-\Gamma$ in the BZ.}
    \label{fig:imp_samp_fitting_6hoppings-1}
\end{figure}

In model {\bf C} the hopping parameters $t_i, i=1,\dots , 6$ are non-zero in Eq.~\eqref{eq:hop-mats}. In this case no eigenfrequency can be computed analytically, and we had to resort to a numerical diagonalization of the Hamiltonian~\eqref{eq:hk} at the points of the BZ. The results of the importance rMC are shown in Fig.~\ref{fig:imp_samp_fitting_6hoppings-1}. The values of the optimal parameters are $t_1=1$ and $(\epsilon, t_2, t_3, t_4, t_5, t_6) = (-5.39703, 0.985327, -0.684177, -0.0983953, 0.0362692,\\ 0.0365651 )$. We observe a much better agreement of the fit with the experimental data as compared to the fits based on models {\bf A} and {\bf B}. Furthermore, in distinction to models {\bf A} and {\bf B}, model {\bf C} provides a good description of the experimental one at the upper band edge of the second band. 

\section{Properties of the eigenstates of the honome-based TBM matrix\label{Comparison}} 
\begin{figure}[h]
\centering
\includegraphics[width=1.0\columnwidth]{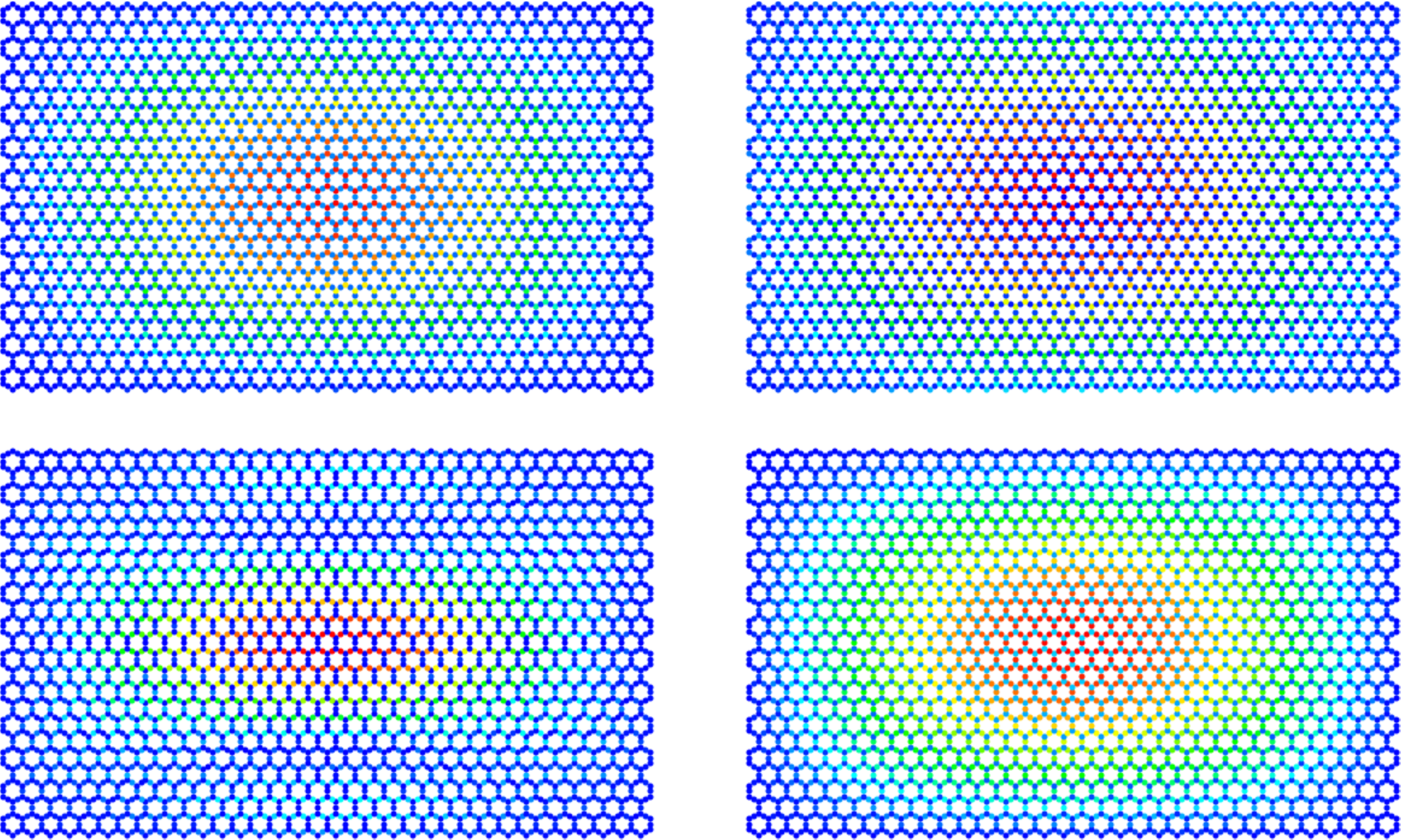}
	\caption{Eigenmodes with eigenfrequencies of the honome TBM matrix emulating the Dirac microwave billiard at the lower and upper edge of the bands framing the lower DP FB (upper panels), the lower edge of the FB (left lower panel) and upper edge of the bands framing the upper DP (right lower panel). Below the FB the eigenmodes are maximal on honeycomb-sublattice sites. In the other examples they are localized on kagome-sublattice sites. The color code is the same as in Fig.~\ref{fig:WFW}.
}
\label{fig:WFH}
\end{figure}

To further corroborate our supposition that the microwave Dirac billiard provides an experimental realization of a honome lattice, we compared distributions of their eigenmodes and their spectral fluctuation properties. The eigenmodes and eigenfrequencies of the honome lattice were obtained by diagonalizing the $(4082\times 4082)$ dimensional TBM matrix, where the sites of the underlying honome and kagome sublattices were chosen at the positions of the centers of the voids and between adjacent metal cylinders of the microwave resonator, respectively. Throughout the section, we use the hopping parameters obtained with model {\bf C}, since the corresponding DOS agreed best with the experimental one. The eigenmodes correspond to the squared eigenvector components associated with the sites of the honome lattice. Figure~\ref{fig:WFH} shows examples of eigenmodes at the lower (left upper panel) and upper (right upper panel) band edges of the region below the FB, at the lower edge of the FB (left lower panel) and at the upper edge of the bands framing the upper DP (right lower panel). In the bands framing the lower DP the wave functions are maximal at the sites of the honeycomb sublattice, in the FB region they are non-vanishing only on kagome sites,~\cite{jacqmin2014direct} whereas in the region above the FB they may be maximal on honeycomb- or kagome-sublattice sites, in accordance with the observations made for the corresponding electric-field intensity distributions of the microwave Dirac billiard.  

\begin{figure}[h]
\centering
\includegraphics[width=1.0\columnwidth]{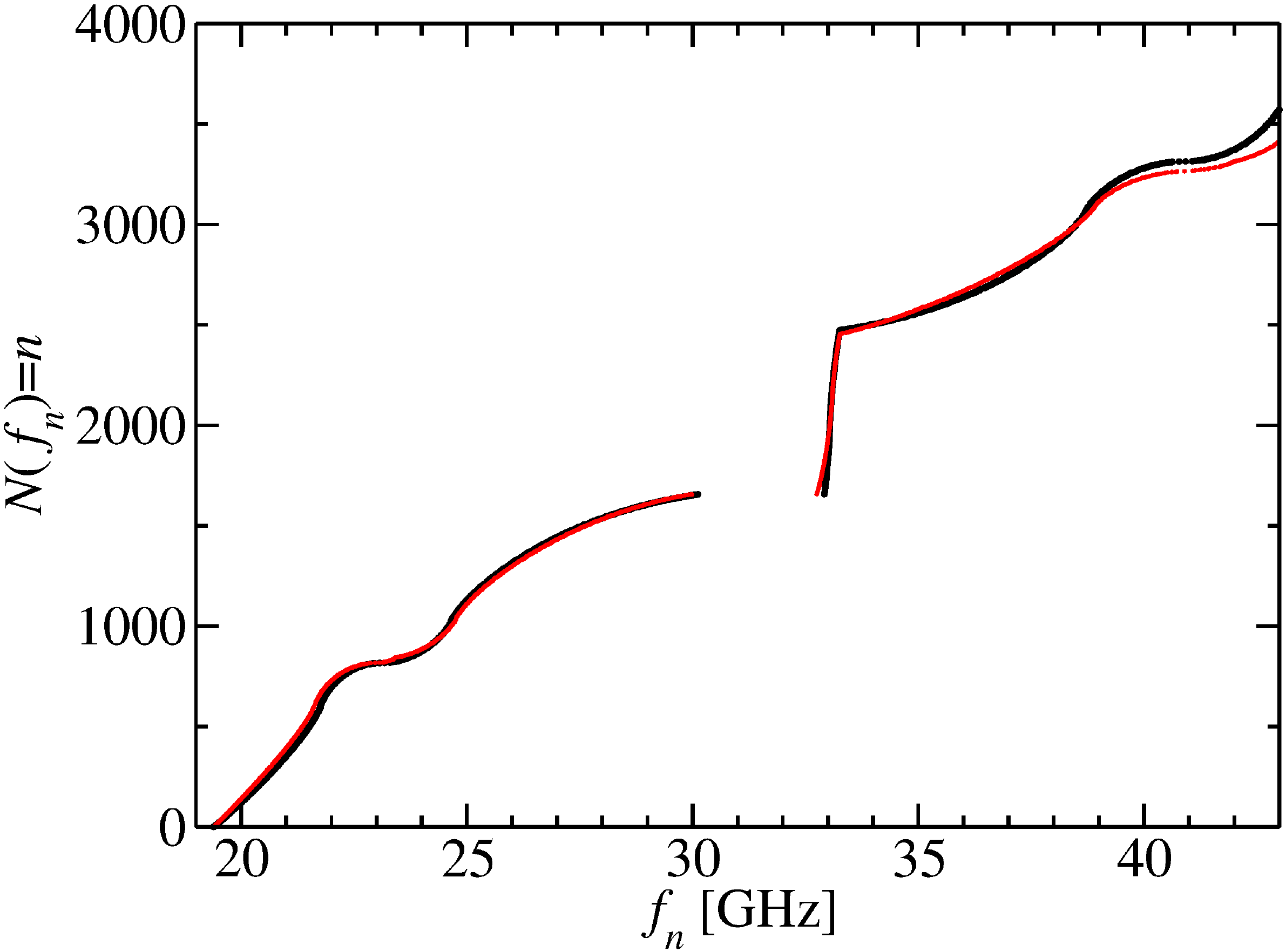}
	\caption{Integrated DOS of the microwave Dirac billiard (black) and the honome TBM matrix (red). The plateaus correspond to the regions of low spectral desnity around the DPs, the steep ascend to the region of the FB and the kinks are due to the van Hove singularities. 
}
\label{fig:NH}
\end{figure}

In Fig.\ref{fig:NH} the integrated DOS, i.e., the number of resonance frequencies $f_n$, which are sorted by increasing value, below $f_n$, $N(f_n)=n$ of the Dirac billiard is compared to that of the honome lattice. The plateaus correspond to the regions of low spectral density around the DP, the steeply ascending part to that of the FB and the van Hove singularities are recognizable as kinks. Deviations between the two curves may be attributed to the fact, that the boundary conditions slightly differ at the zigzag edges, i.e., the longer sides of the rectangle, since for the Dirac billiard the wave functions vanish at the walls of the rectangular basin, whereas for the honome lattice they vanish on the sites along the first outer row bordering it. As a consequence, the number of eigenstates in the flatband differ, thus explaining the deviation of about $\delta_n=50$ between the two curves above the FB. 
In the following subsections we present results for the eigenmode distributions and the spectral properties of the honome lattice and compare them to those of the microwave Dirac billiard.
\subsection{Eigenmode properties\label{WF}}
\begin{figure*}[ht]
\centering
\includegraphics[width=17cm]{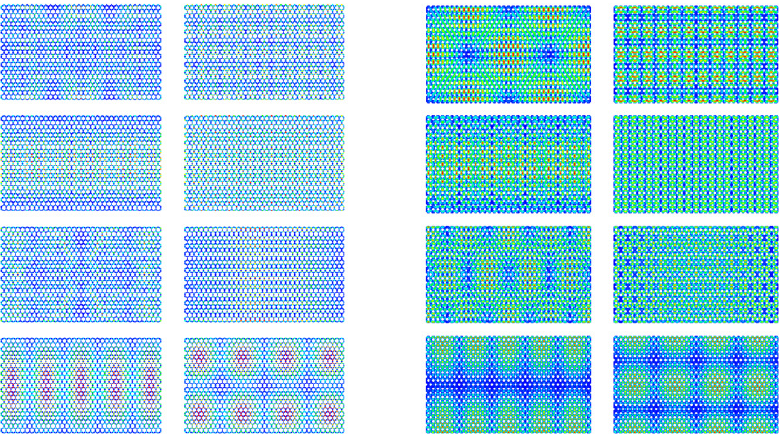}
	\caption{Eigenmodes with eigenfrequencies of the honome TBM matrix (left part) emulating the microwave Dirac billiard and electric-field intensity distributions (right part) in the microwave Dirac billiard corresponding to resonance frequencies in the bands framing the lower DP. Examples are shown for the regions around the lower van Hove singularity (1st rows), the DP (2nd rows), the upper van Hove singularity (3rd rows) and the upper band edge (4th rows). The color code is the same as in Fig.~\ref{fig:WFW}.
}
\label{fig:WF1}
\end{figure*}

\begin{figure*}[ht!]
\centering
\includegraphics[width=17cm]{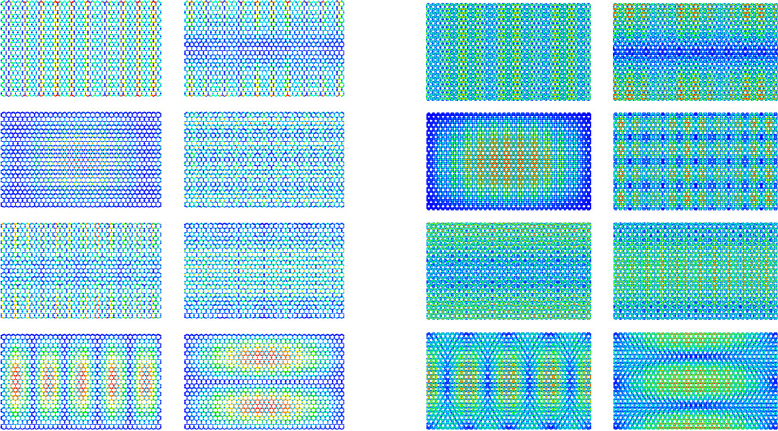}
\caption{Eigenmodes with eigenfrequencies of the honome TBM matrix (left part) emulating the microwave Dirac billiard and electric-field intensity distributions (right part) in the microwave Dirac billiard corresponding to resonance frequencies in the bands framing the upper DP. Examples are shown for the regions close to the lower band edge (1st rows), the lower van Hove singularity (2nd rows), the DP (3rd rows) and the upper band edge (4th rows). The color code is the same as in Fig.~\ref{fig:WFW}.
}
\label{fig:WF2}
\end{figure*}

\begin{figure*}[ht!]
\centering
\includegraphics[width=17cm]{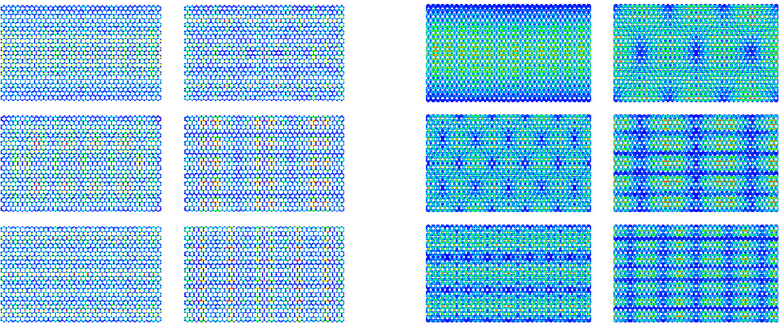}
\caption{Eigenmodes with eigenfrequencies of the honome TBM matrix (left part)  emulating the microwave Dirac billiard and electric-field intensity distributions (right part) in the microwave Dirac billiard corresponding to resonance frequencies of the microwave Dirac billiard in the FB. The color code is the same as in Fig.~\ref{fig:WFW}.
}
\label{fig:WF3}
\end{figure*}

In Ref.~\onlinecite{Bittner2012} electric-field intensity distributions were measured in the region of low resonance density, that is, around the DP with a microwave Dirac billiard containing 273 metal cylinder. This was not possible~\cite{dietz2015spectral} in the present microwave Dirac billiard, because of the large number of closely lying metallic cylinders. Therefore, we compare the squared-eigenmode distributions of the honome lattice to electric-field distributions computed from the Helmholtz equation, of which examples are shown in Fig.~\ref{fig:WFW}. We present a few typical distributions in Figs.~\ref{fig:WF1} and~\ref{fig:WF2} in the bands framing the lower and the upper Dirac point, respectively, and in Fig.~\ref{fig:WF3} in the region of the flatband. Shown are from up to down in Fig.~\ref{fig:WF1} examples around the lower van Hove singularity (1st rows), the DP (2nd rows), the upper van Hove singularity (3rd rows) and the upper band edge (4th rows) and in Fig~\ref{fig:WF2} for the regions close to the lower band edge (1st rows), the lower van Hove singularity (2nd rows), the DP (2nd) and the upper band edge (4th rows). 
For both systems the first $~200-250$ eigenmodes starting from the lower or the upper band edge of the region below the FB or from the upper one above the FB resemble those of the corresponding rectangular quantum billiard.~\cite{dietz2015spectral} Thus, in these regions the honome lattice is effectively described by the Hamiltonian of the quantum billiard, which is in agreement with the observations made for the microwave Dirac billiard. Note, that in some cases the spatial patterns of the eigenmodes of the TBM matrix correpond to those of a quantum billiard with Dirichlet boundary conditions, like those shown in the 4th row of the left part of Figs.~\ref{fig:WF1} and~\ref{fig:WF2}, whereas for the microwave Dirac billiard they coincide with those of a quantum billiard with Neumann boundary conditions, shown in the 4th row of the right part of these figures.  This may be attributed to their slightly differing boundary conditions. An analogy with the corresponding quantum billiard, however, is not found at the lower band edge of the region above the FB. There, the eigenmodes clearly differ from those of the corresponding rectangular quantum billiard. Examples are shown in the top rows of Fig.~\ref{fig:WF2}. Note, that the eigenmodes of the honeycomb-based TBM resemble the wave functions of the quantum billiard in that region, and thus this TBM fails there to describe the electric field distributions, though the spectral properties were shown to be similar in Ref.~\onlinecite{dietz2015spectral}. Close to the DPs the features of the eigenmode distributions resemble those of the corresponding graphene billiard. At the DPs, the eigenmodes are vanishingly small, except at zigzag edges.~\cite{Wurm2011} Therefore, we do not show such examples. Furthermore, we observe that in the bands framing the lower DP the eigenmodes are maximal at sites of the honeycomb sublattice and may also be large at kagome-sublattice sites, implying that there the hopping between honycomb-sublattice sites via the kagome sublattice sites is strong, that is, the wave-function overlap is large. In the bands framing the upper DP the eigenmodes can be large at the sites of the honeycomb and the kagome sublattice, whereas in the FB all eigenmodes are vanishingly small on the honeycomb-sublattice sites.~\cite{jacqmin2014direct} This is expected, since the nearly FB is a remnant of the perfect FB of model {\bf A} that only has support on kagome-sublattice sites. These observations are in accordance with those made for the electric-field intensity distributions of the microwave Dirac billiard. The applicability of the honome-based TBM is furthermore confirmed by the spectral properties.
\subsection{Spectral properties\label{Spectral}}
\begin{figure}[ht!]
\centering
\includegraphics[width=0.7\columnwidth]{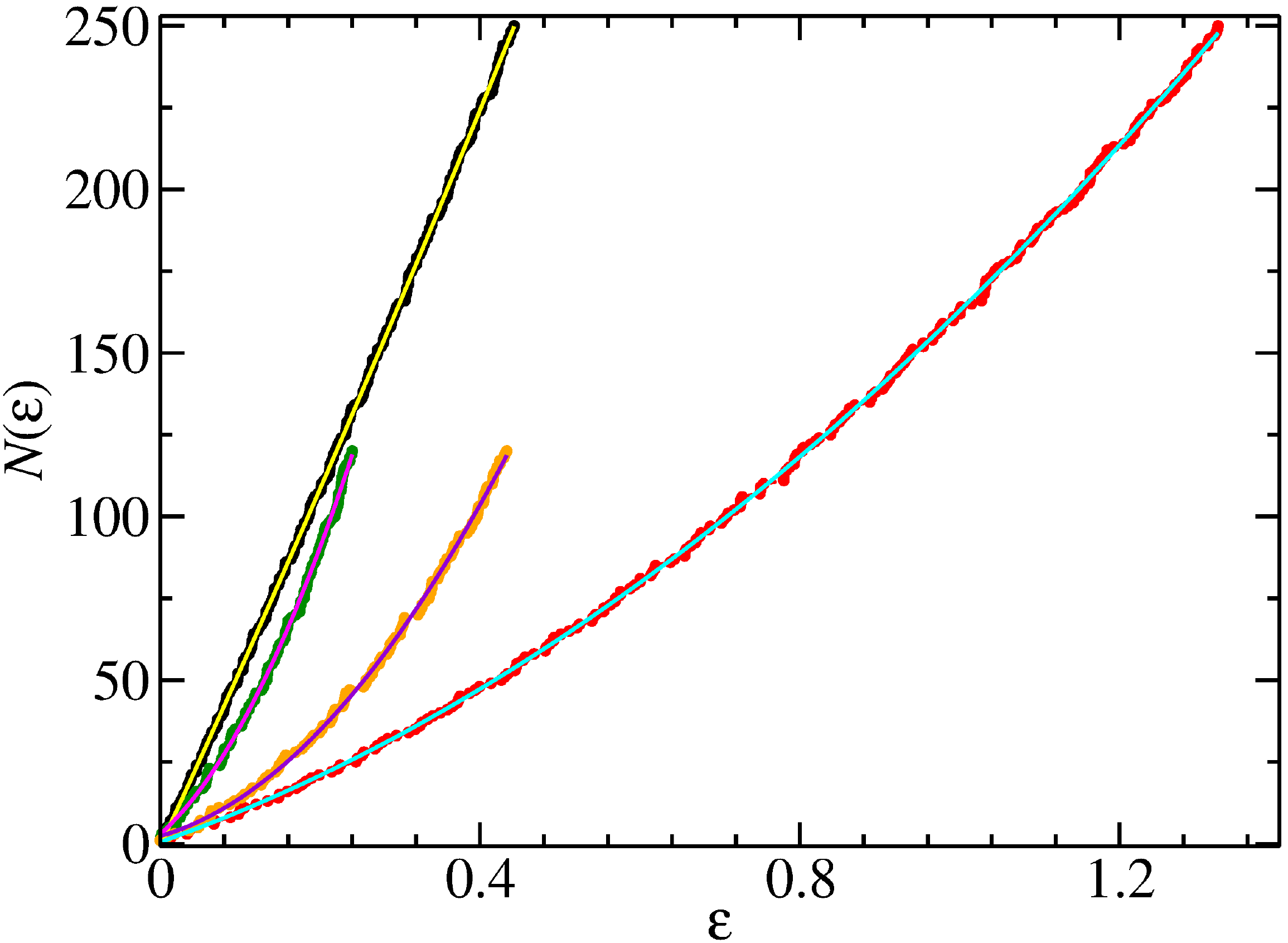}
\caption{Integrated DOS of the first 250 eigenfrequencies of the honome TBM matrix starting from the lower (black dots) and upper (red dots) band edges denoted $f_{BE}$ ($\epsilon =\vert f-f_{BE}\vert$) and the first 100 eigenfrequencies starting from the DP denoted $f_D$ ($\epsilon =\vert f-f_{DP}\vert$) below (green dots) and above (orange dots) the DP in the region below the FB. The yellow, cyan, magenta and violet curves exhibit the best-fitting second-order polynomials to the respective integrated DOS. 
}
\label{fig:SP1}
\end{figure}

We, furthermore, investigated fluctuation properties in the eigenfrequency spectrum of the honome TBM matrix, using the hopping parameters from model {\bf C}, and compared them to random-matrix theory predictions for quantum systems with an integrable or chaotic classical dynamics. These were conjectured to be universal and to coincide with those  of uncorrelated Poissonian random numbers, and of the eigenvalues of random matrices from the Gaussian orthogonal ensemble (GOE), respectively.~\cite{Berry1977,Casati1980,Bohigas1984} Examples for an integrable and a chaotic billiard are the rectangular and the Sinai billiard, which is constructed by inserting a hard-wall disk into a rectangular billiard, respectively.~\cite{stoeckmann1990quantum} To obtain information on the universality of the spectral fluctuation properties, first system-specific properties need to be extracted. This is done by replacing the eigenfrequencies $\epsilon_n$ by the smooth part of the integrated DOS $N(\epsilon_n)$, $\tilde\epsilon_n=N^{smooth}(\epsilon_n)$. However, for this an analytical expression is needed for $N^{smooth}(\epsilon_n)$. Starting from the band edges at $f_{BE}$ and defining $\epsilon_n=\vert f_n-f_{BE}\vert$ or from the Dirac points at $f_{DP}$, $\epsilon_n=\vert f_n-f_{DP}\vert$, the corresponding $N^{smooth}(\epsilon_n)$ is well approximated by a second-order polynomial, as illustrated in Fig.~\ref{fig:SP1} for the region below the FB.

\begin{figure}[ht!]
\centering
\includegraphics[width=0.9\columnwidth]{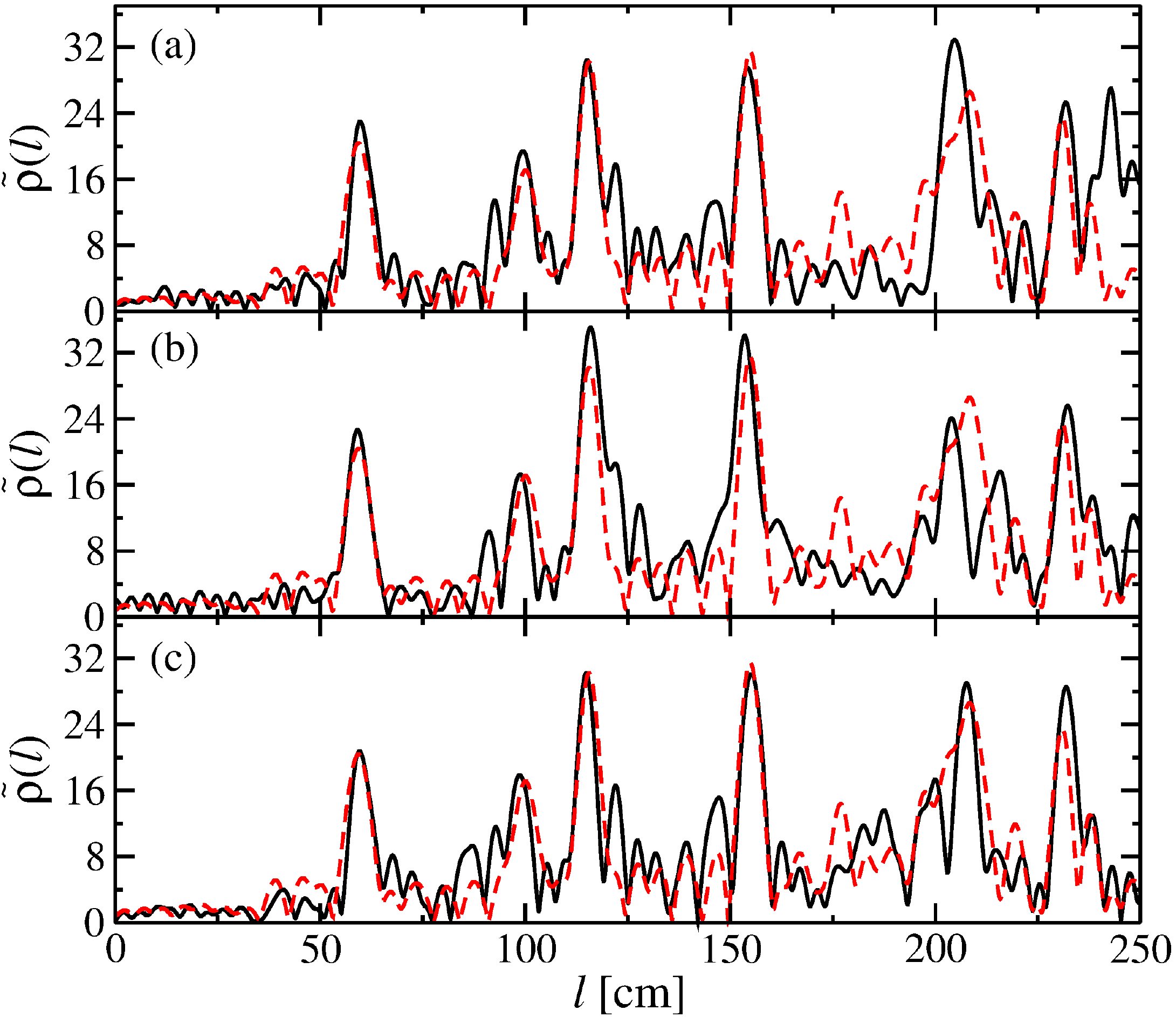}
        \caption{Length spectra obtained from the first 250 eigenfrequencies of the honome TBM matrix starting from the lower (black dots) and upper (red dots) band edges below the FB and the upper one above the FB.
}
\label{fig:SP0}
\end{figure}

To verify, whether the analogy between the eigenmodes of the TBM matrix and those of the corresponding quantum billiard, observed in the eigenfrequency ranges near the band edges, also holds for the spectral properties we first computed length spectra, that is, the modulus of the Fourier transform of the fluctuating part of the DOS $\rho^{fluc}(k)=\frac{{\rm d}[N(k)-N^{smooth}(k)]}{{\rm d} k}$ from wave numbers $k_n=\frac{2\pi f_n}{c}\leq k_{max}, n=1,2,\dots$ to lengths l, $\tilde\rho(l)=\vert\int_0^{k_{max}}{\rm d}ke^{ikl}\rho^{fluc}(k)\vert$. In Fig.~\ref{fig:SP0} we compare length spectra obtained from the first 250 eigenfrequencies starting from the lower [(a)] and the upper [(b)] band edges below the FB and from the upper one [(c)] above the FB with that of the rectangular quantum billiard with Dirichlet boundary conditions. Here, we rescaled the eigenfrequencies of the TBM matrix as described in Ref.~\onlinecite{dietz2015spectral} using their functional relation to the eigenvalues of the quantum billiard. The length spectra indeed exhibit peaks at the lengths of the periodic orbits of the corresponding classical billiard, thus confirming the analogy between the honome lattice and the quantum billiard also for the eigenfrequencies. These results agree very well with those obtained in Ref.~\onlinecite{dietz2015spectral} for the experimental eigenfrequencies.   

\begin{figure}[ht!]
\centering
\includegraphics[width=\columnwidth]{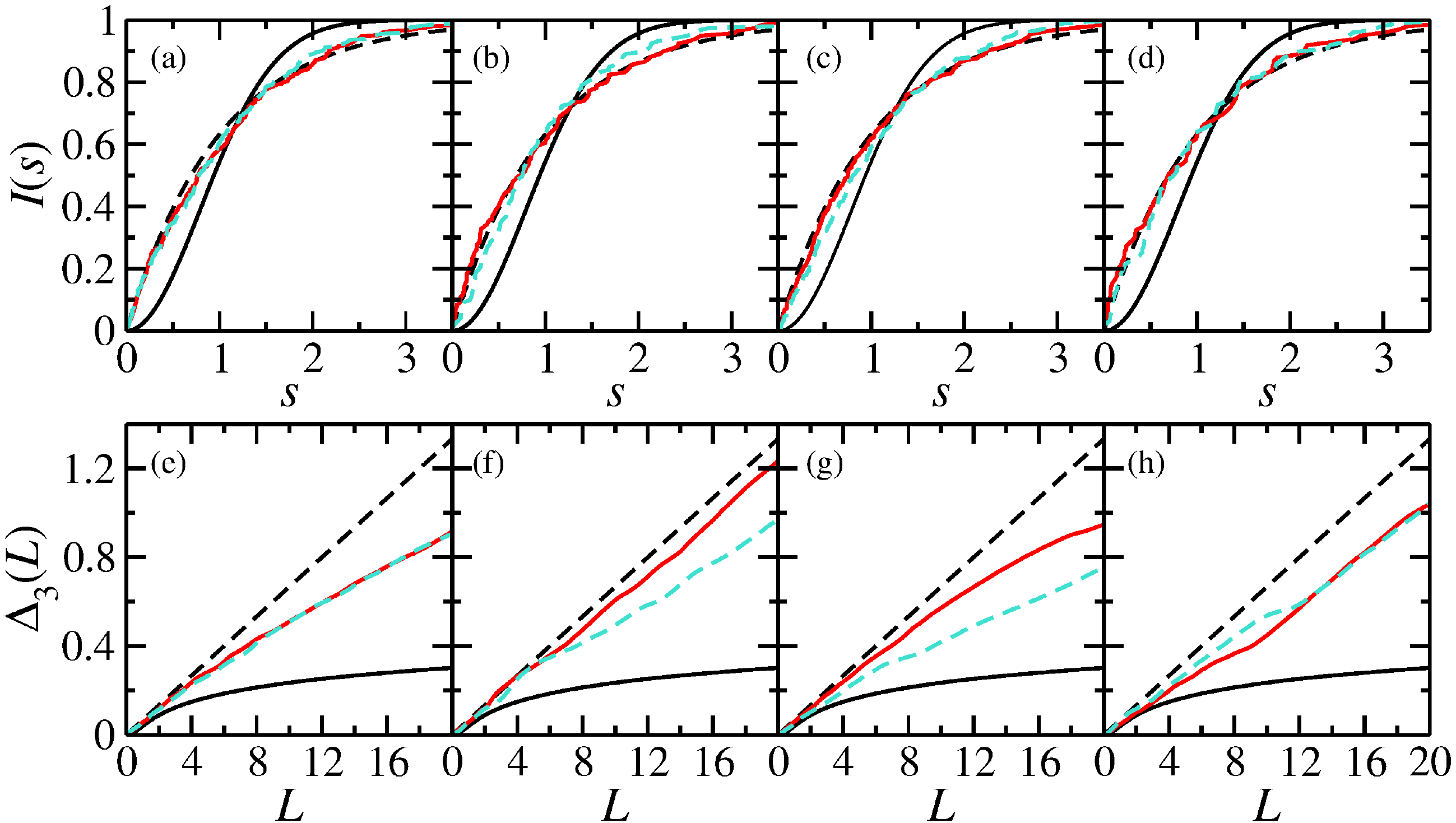}
	\caption{Cumulative nearest-neighbor spacing distributions (upper panels) and $\Delta_3$ statistics (lower panels) for the honome TBM below [(a),(b),(e),(f)] and above [(c),(d),(g),(h)] the FB, using the first 250 eigenfrequencies starting from the lower (red full lines) and upper (turquoise dashed lines) band edges [(a),(e)] and [(c),(g)], and the first 100 eigenfrequencies starting from the DP below (red full lines) and above (turquoise dashed lines) it [(b),(f)] and [(d),(h)]. The full and dashed black lines depict the corresponding GOE and Poisson curves.
}
\label{fig:SP2}
\end{figure}

Proceeding as in Ref.~\onlinecite{dietz2015spectral}, we, furthermore, investigated the nearest-neighbor spacing distribution and its cumulative distribution, the number variance and the $\Delta_3(L)$ statistics, which provides the mean least-squares deviation of the integrated DOS from the straight line best fitting it in an interval of length $L$, in the regions around the band edges and around the DPs below and above the FB. Figure~\ref{fig:SP2} shows results for the cumulative nearest-neighbor spacing distribution and the $\Delta_3$ statistics. In all considered frequency regions we find good agreement with Poisson statistics (dashed black lines). Note, that the deviations observed for the $\Delta_3$ statistics are of the same size as for the microwave Dirac billiard~\cite{dietz2015spectral} and are attributed to the fact that the ratio of the side lengths of the rectangular billiard $R=420/249.4$ is too close to a rational number. This drawback becomes visible especially in the long-range correlations quantified, e.g., by the $\Delta_3$ statistics. For the matter of completeness we also include the corresponding GOE curves. 

\begin{figure}[ht!]
\centering
\includegraphics[width=\columnwidth]{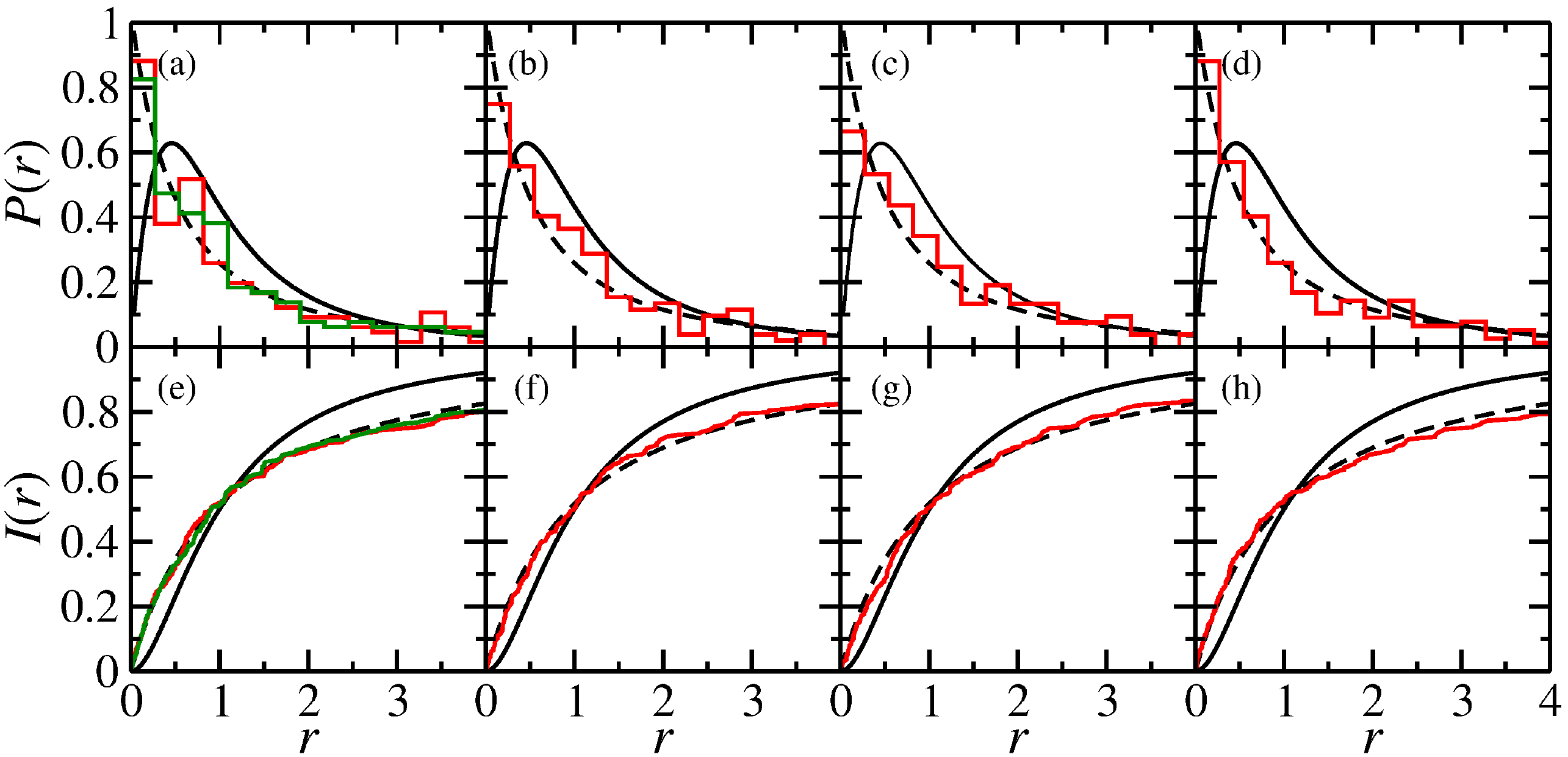}
\includegraphics[width=\columnwidth]{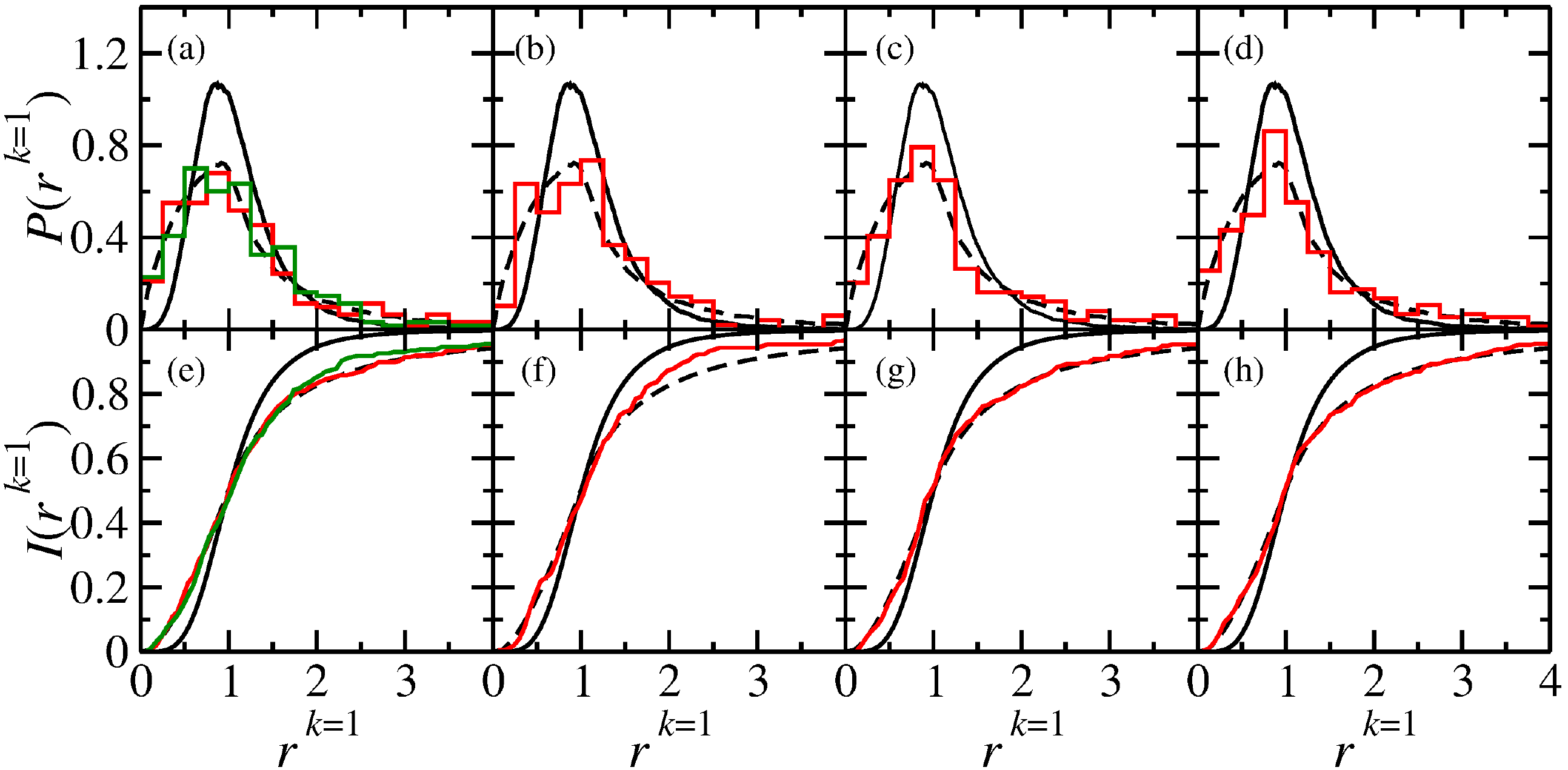}
	\caption{Upper part: Ratio distributions [(a)-(d)] and cumulative ratio distributions [(e)-(h)]  for the honome TBM below the FB for the first 250 eigenfrequencies starting from the lower (red) and upper (green) band edges [(a),(e)] and 200 eigenfrequencies around the lower van Hove singularity [(b),(f)], the upper van Hove singularity [(c),(g)] and the DP [(d),(h)]. The full and dashed black lines depict the corresponding GOE and Poisson curves.
	Lower part: Same as upper part for the (k=1)-overlapping ratio distributions and their cumulative distributions.
}
\label{fig:SP3}
\end{figure}

\begin{figure}[ht!]
\centering
\includegraphics[width=\columnwidth]{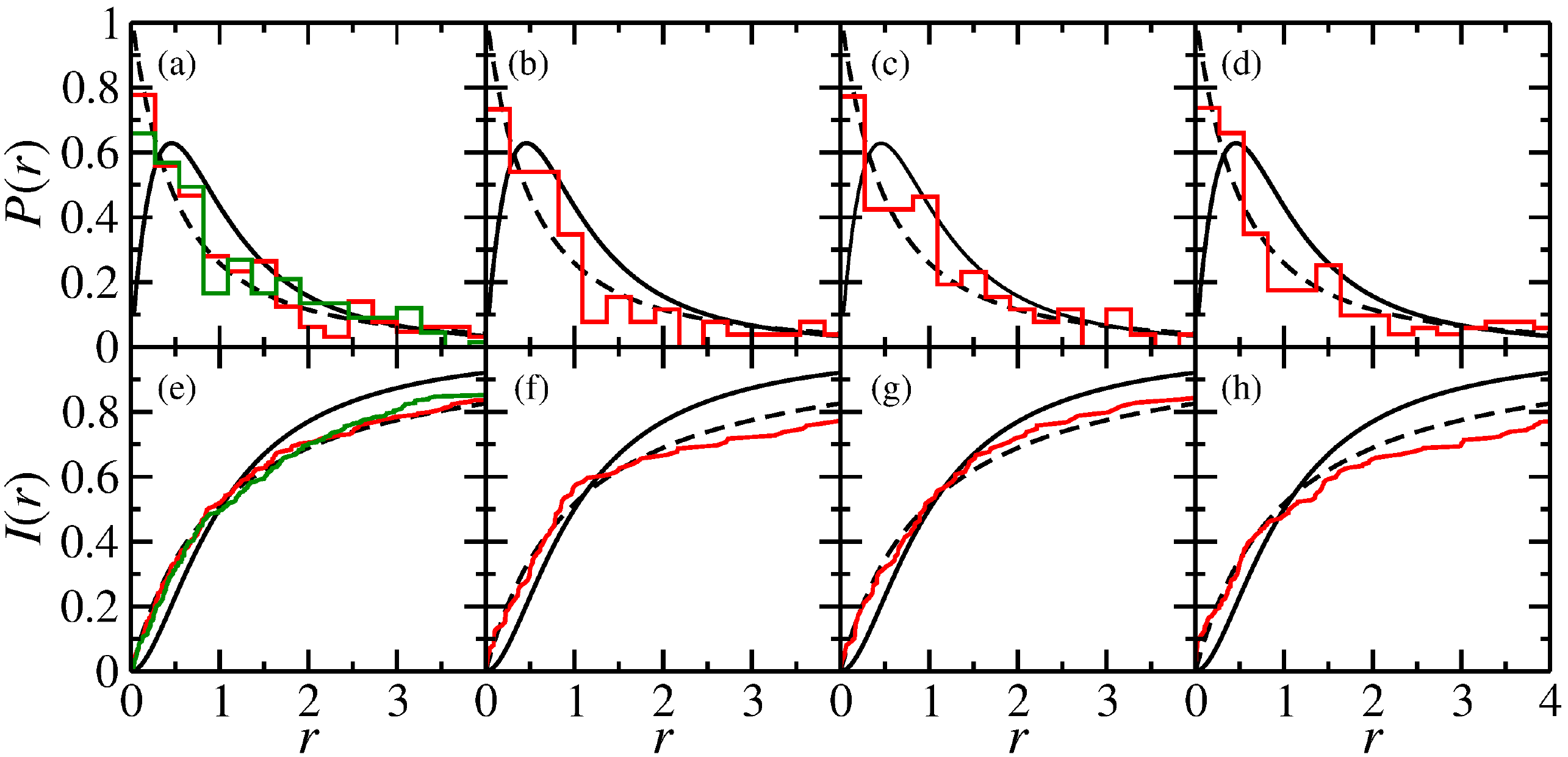}
\includegraphics[width=\columnwidth]{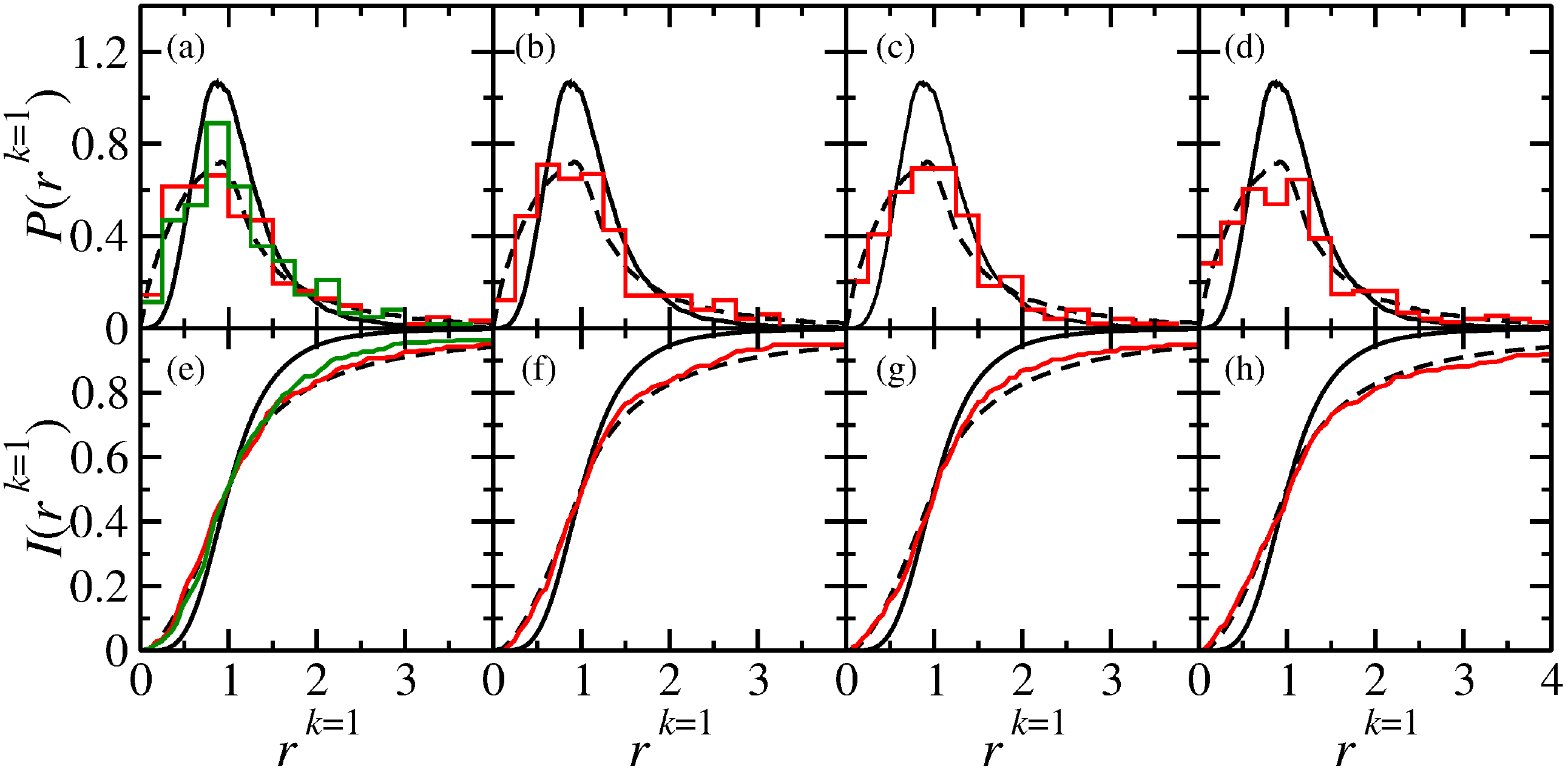}
	\caption{Same as Fig.~\ref{fig:SP3} in the region above the FB.
}
\label{fig:SP4}
\end{figure}

In the vicinity of the van Hove singularities such a procedure of unfolding is not possible, as there the DOS diverges logarithmically with the number of sites~\cite{Dietz2013}. Yet, unfolding is not needed when considering the distribution of the ratios~\cite{Oganesyan2007,Atas2013} $r_i=\frac{\epsilon_{i+1}-\epsilon_i}{\epsilon_{i}-\epsilon_{i-1}}$ and the $k$th overlapping ratio distribution~\cite{Atas2013a} of $r^k_i=\frac{\epsilon_{i+k+1}-\epsilon_i}{\epsilon_{i+k}-\epsilon_{i-1}}$ which are dimensionless and thus do not depend on unfolding as long as the DOS does not vary on the scale of the mean spacing. It was demonstrated in Ref.~\onlinecite{Dietz2016}, that they are applicable to the regions of van Hove singularities. We found good agreement with Poisson statistics in all frequency ranges below and above the FB, thus recovering the results of Ref.~\onlinecite{Dietz2016}, shown in Figs.~\ref{fig:SP3} and~\ref{fig:SP4}. 

\begin{figure}[ht!]
\centering
\includegraphics[width=0.7\columnwidth]{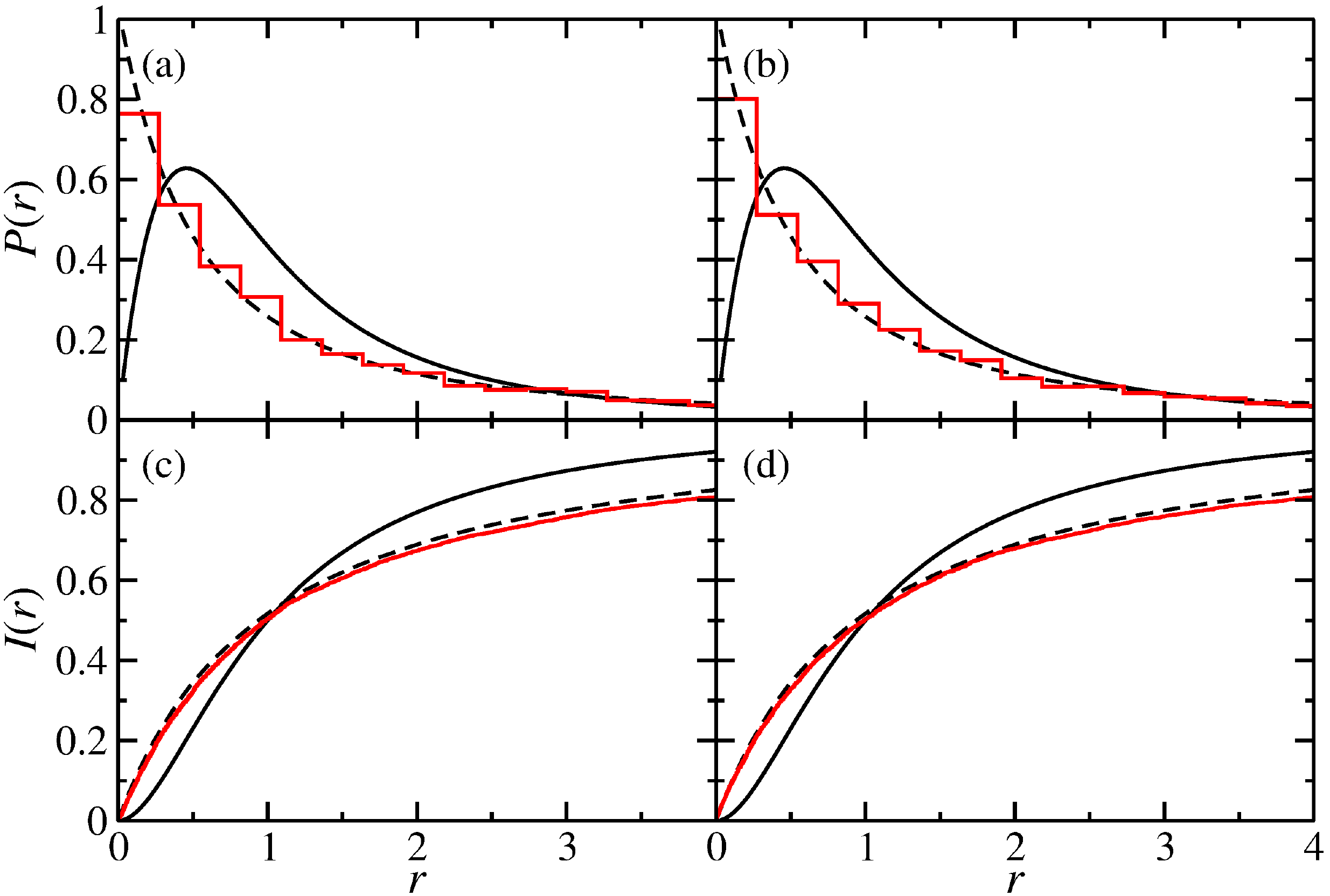}
	\caption{Ratio distributions [(a),(b)] and cumulative ratio distributions of all resonance frequencies of the microwave Dirac billiard [(a),(c)] and all eigenfrequencies of the honome TBM matrix [(b),(d)]. The full and dashed black lines depict the corresponding GOE and Poisson curves.
}
\label{fig:SP5}
\end{figure}

\begin{figure}[ht!]
\centering
\includegraphics[width=0.7\columnwidth]{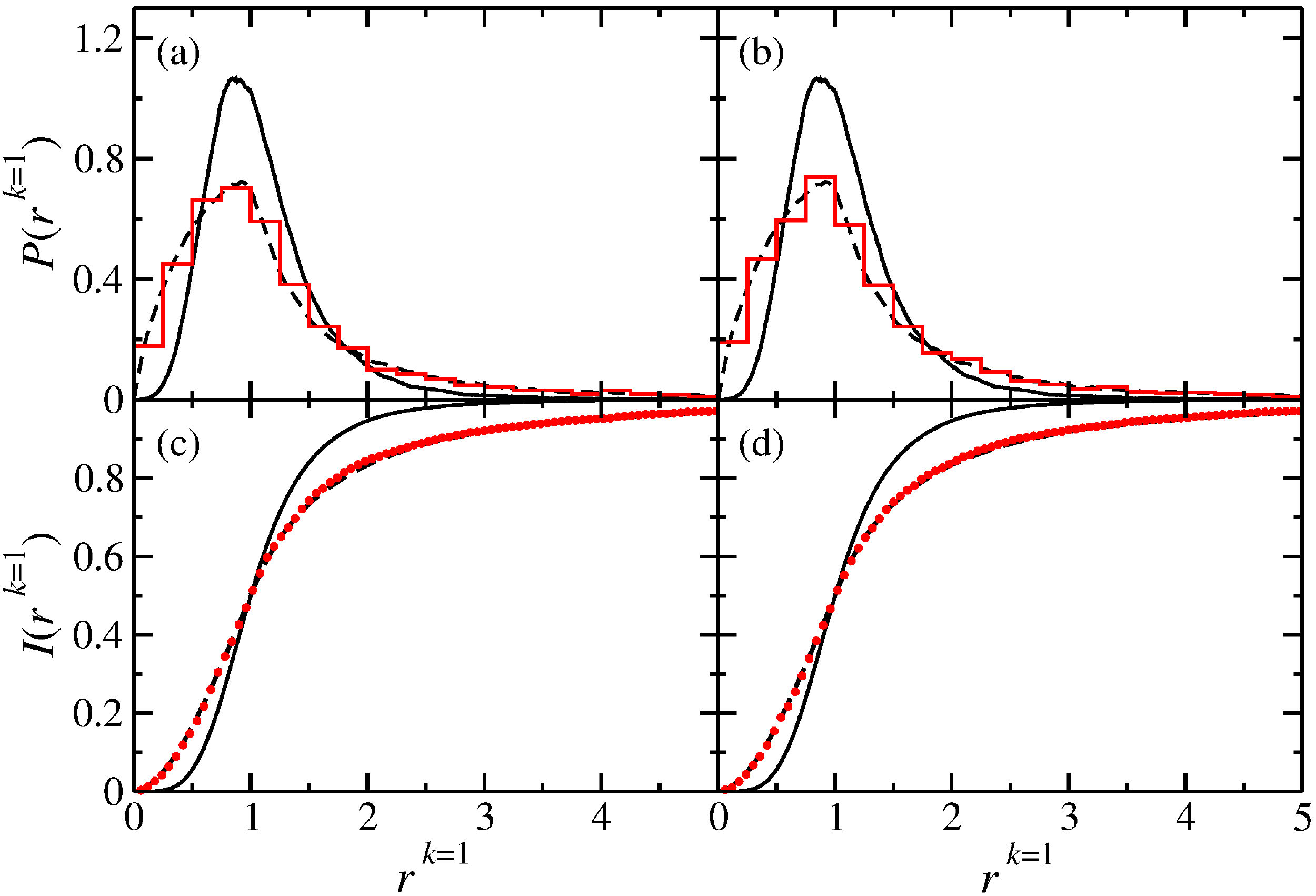}
\caption{Same as Fig.~\ref{fig:SP5} for the (k=1)-overlapping ratio distributions and their cumulative distributions.
}
\label{fig:SP6}
\end{figure}

\begin{figure}[ht!]
\centering
\includegraphics[width=0.7\columnwidth]{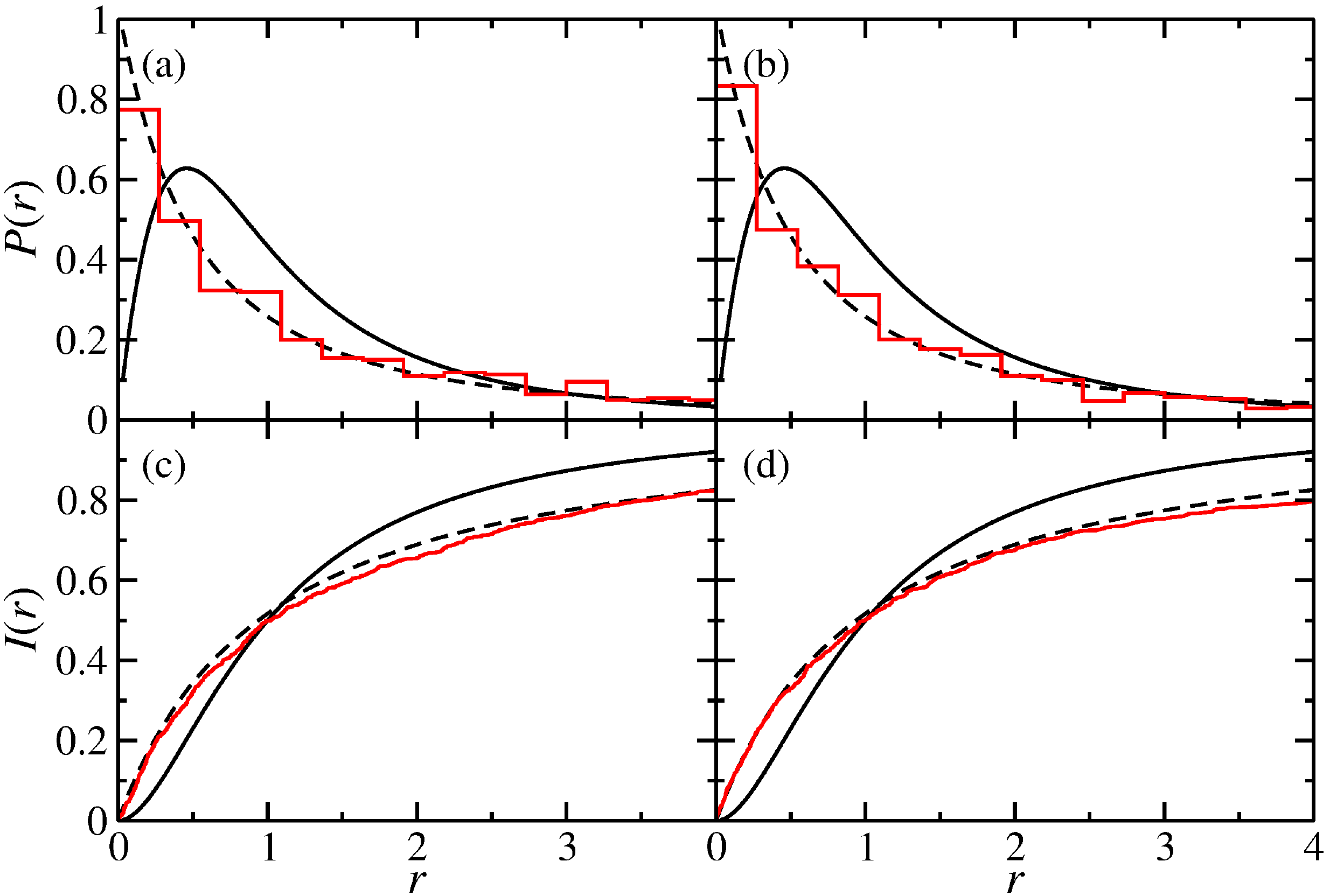}
\caption{Ratio distributions [(a),(b)] and cumulative ratio distributions of the resonance frequencies of the microwave Dirac billiard [(a),(c)] and the eigenfrequencies of the honome TBM matrix [(b),(d)] in the region of the FB. The full and dashed black lines depict the corresponding GOE and Poisson curves.
}
\label{fig:SP7}
\end{figure}

\begin{figure}[ht!]
\includegraphics[width=0.7\columnwidth]{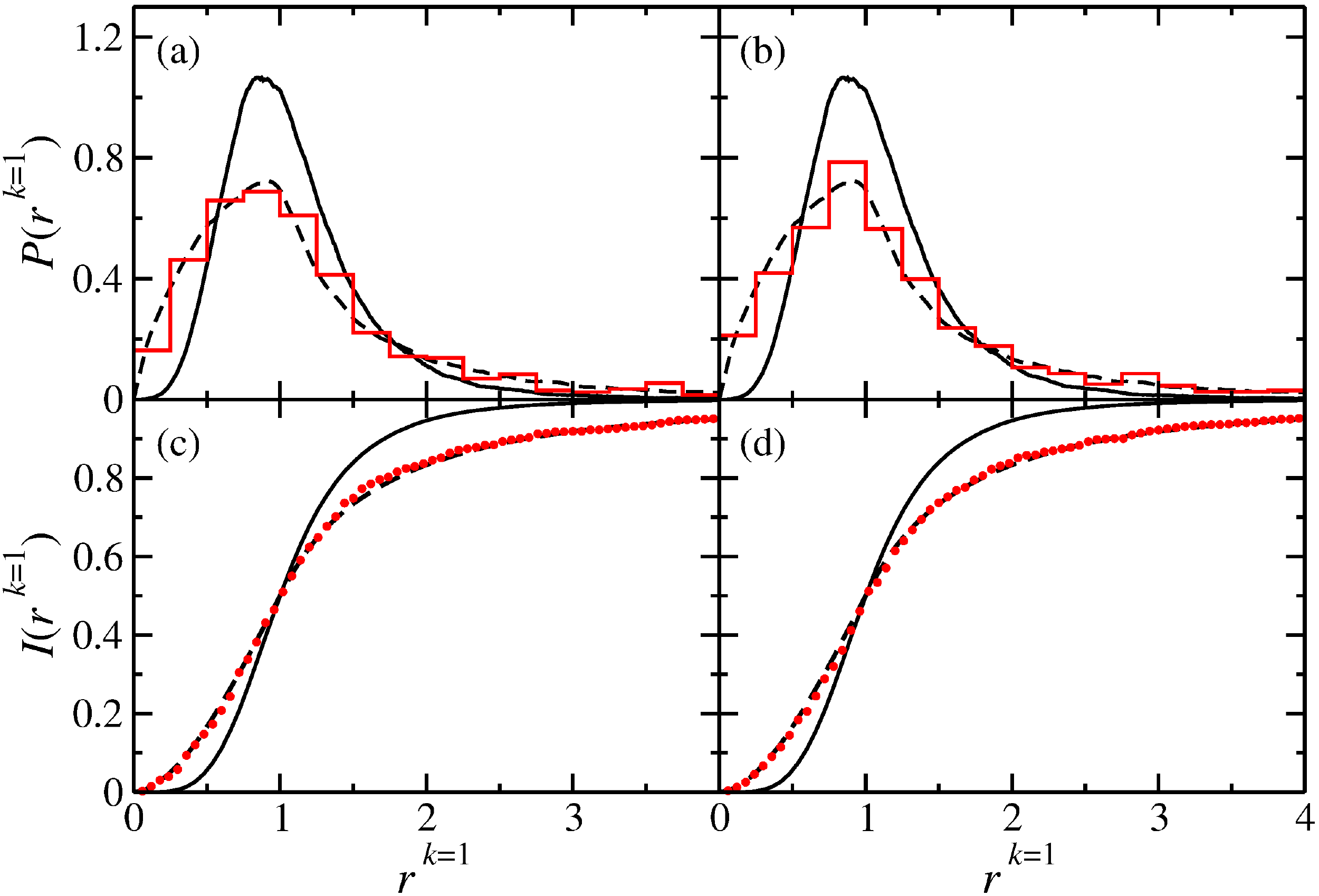}
\caption{Same as Fig.~\ref{fig:SP7} for the (k=1)-overlapping ratio distributions and their cumulative distributions.
}
\label{fig:SP8}
\end{figure}

We furthermore evaluated the distributions taking into account all eigenfrequencies and those in the FB, respectively, and compared them to those of the resonance frequencies of the microwave Dirac billiard. Here, we considered the experimentally and the numerically determined resonance frequencies, as in that region about $10\%$ of the resonances could not be resolved in the experiments because they are partly too closely lying, however, found in both cases good agreement with Poisson as illustrated in Figs.~\ref{fig:SP5}-~\ref{fig:SP8}.   
\section{Conclusions and discussions
\label{sec:conclusion}}

We propose a TBM which is based on a honome lattice, consisting of a honeycomb and a kagome sublattice, as a model to describe the properties of a flat microwave photonic crystal containing metal cylinders which are arranged on a triangular lattice. This supposition was motivated by the fact that the electric-field intensity inside the resonator is either maximal at the voids formed by, respectively, three cylinders, that is, localized on the sites of a honeycomb-lattice structure or at the border of the voids, at the centers between adjacent metal cylinders which constitute a kagome-lattice structure. To be more explicit, we found that below the FB the electric-field intensity is maximal at the voids and might be nonvanishing between adjacent cylinders, thus indicating an overlap between the electric field modes centered at neighboring voids. In the region of the FB the electric-field intensity is localized between adjacent cylinders, and thus exhibits kagome-lattice type spatial patterns. Finally, above the FB it can be localized at the voids or between adjacent cylinders. Agreement between the experimental and TBM DOS is good, when including up to at least 6th n.n. hoppings on the honome lattice corresponding to 2nd n.n. and 3rd n.n. in the honeycomb and kagome sublattices, respectively. We also performed TBM computations with up to 8th n.n. hopping in the honome lattice corresponding to 3rd n.n. hopping in the honeycomb-lattice based TBM yielding a barely visible improvement. This extension of the parameter space implies that an additional hopping matrix needs to be added to Eq.~\eqref{eq:hop-mats} which corresponds to hoppings between n.n. unit cells in the vertical direction $\ve_4=\ve_1+2\ve_2$. To determine the set of parameters which yields the best honome-based TBM description of the experimental DOS, we developed a numerical algorithm based on the reverse Monte-Carlo method which identifies the best fit of the TBM DOS to the experimental one. 

In order to corroborate our surmise, that the appearance of two DPs together with a FB in the resonance frequency spectrum and DOS of microwave Dirac billiards may be interpreted as the coupling between atoms located on a honeycomb and a kagome sublattice, respectively, that is, may be described by a honome-based TBM, we compared their eigenmode distributions and spectral properties and found good agreement. Furthermore, we demonstrate, that in the vicinity of the band edges below the FB and of the upper band edge above the FB the eigenstates of the honome TBM matrix are described by an effective Hamiltonian which coincides with that of a quantum billiard of corresponding shape with either Dirichlet or Neumann boundary conditions. We may conclude from our findings that the appearence of a FB together with a second DP framed by van Hove singularities may be attributed to the overlap between the electric-field components centered on the honeycomb and kagome sublattices and thereby give an answer to the longstanding question concerning the origin of the FB. Note, that the lower DP and the bands framing it are well separated from the FB and the upper DP, and there the electric-field modes are maximal on the honeycomb sites. This explains why below the FB their features are also well described by a honeycomb-based TBM which, in addition to the hopping parameters comprises overlap parameters accounting for the overlap of the eigenmodes centered at the voids, so that the microwave photonic crystal provides an experimental realization of artificial graphene and of honome-lattice structured materials. We, indeed, provide evidence that such devices serve as a suitable testbed for the experimental investigaton of the features of honome lattice structures with high precision.      

\begin{acknowledgments}
    This work was supported by the Institute for Basic Science in Korea (IBS-R024-D1). We thank Sergej Flach, who initiated this work. BD thanks the Institute of Basic Science at the Center of Theoretical Physics of Complex Systems in Daejeon, Republic of Korea for their hospitality and the National Natural Science Foundation of China for financial support under Grant Nos. 11775100 and 11961131009. 
\end{acknowledgments}

\bibliography{general,flatband}

\end{document}